\documentclass[onefignum,onetabnum]{siamart220329}
\pdfoutput=1

\usepackage{lipsum}
\usepackage{amsfonts}
\usepackage{graphicx}
\usepackage{epstopdf}
\usepackage{algorithmic}
\ifpdf
  \DeclareGraphicsExtensions{.eps,.pdf,.png,.jpg}
\else
  \DeclareGraphicsExtensions{.eps}
\fi

\usepackage{xcolor}
\usepackage{graphicx}
\usepackage{dcolumn}
\usepackage{bm}
\usepackage{diagbox}

\usepackage{siunitx}
\usepackage{subcaption}
\usepackage{multirow}
\usepackage{braket}
\usepackage{tabularx}
\usepackage{booktabs}
\usepackage{comment}

\DeclareSIUnit\angstrom{\text {Å}}
\renewcommand{\vec}[1]{\bm{#1}}
\newcommand{\mat}[1]{\bm{#1}}

\newcommand{\realsd}[1]{\mathbb{R}^{#1}}

\newcommand{\des}[1]{x^{(#1)}}

\newcommand*{\tran}{^{\mkern-1.5mu\mathsf{T}}}
\newcommand{\norm}[2][2]{\left|\left|#2\right|\right|_{#1}}

\newcommand{\inv}{^{\raisebox{.2ex}{$\scriptscriptstyle-1$}}}

\newcommand{\mean}[1]{\overline{#1}}
\newcommand{\diff}[1]{\Delta #1}

\newsiamremark{remark}{Remark}
\newsiamremark{hypothesis}{Hypothesis}
\crefname{hypothesis}{Hypothesis}{Hypotheses}
\newsiamthm{claim}{Claim}

\headers{Computing enthalpies through explainable machine learning}{E. Di Napoli, X. Wu, T. Bornhake, P. Kowalski}

\title{Computing formation enthalpies through an explainable machine learning method: the case of Lanthanide Orthophosphates solid solutions\thanks{Submitted to the editors DATE.
}}

\author{Edoardo Di Napoli\thanks{J\"ulich Supercomputing Centre (JSC), Forschungszentrum J\"ulich GmbH, J\"ulich, Germany 
  (\email{e.di.napoli@fz-juelich.de}).} \thanks{J\"ulich Aachen Research Alliance Center for Simulation and Data Science (JARA-CSD), J\"ulich, Germany,}
\and Xinzhe Wu\thanks{J\"ulich Supercomputing Centre (JSC), Forschungszentrum J\"ulich GmbH, J\"ulich, Germany.}
\and Thomas Bornhake\thanks{Physics Department, RWTH Aachen University, Aachen, Germany.}
\and Piotr M. Kowalski\thanks{Institute of Energy and Climate Research (IEK-13), Forschungszentrum J\"ulich GmbH, J\"ulich, Germany.} \thanks{J\"ulich Aachen Research Alliance JARA Energy & Center for Simulation and Data Science (CSD), J\"ulich, Germany.}}

\usepackage{amsopn}

\ifpdf
\hypersetup{
  pdftitle={Computing formation enthalpies through an explainable machine learning method: the case of Lanthanide Orthophosphates solid solutions},
  pdfauthor={E. Di Napoli, X. Wu, T. Bornhake and P.M. Kowalski}
}
\fi

\begin{document}

\maketitle

\begin{abstract}
In the last decade, the use of Machine and Deep Learning (MDL) methods
in Condensed Matter physics has seen a steep increase in the number of
problems tackled and methods employed. A number of distinct MDL
approaches have been employed in many different topics; from
prediction of materials properties to computation of Density
Functional Theory potentials and inter-atomic force fields. In many
cases the result is a surrogate model which returns promising
predictions but is opaque on the inner mechanisms  of its success. On
the other hand, the typical practitioner looks for answers that are
explainable and provide a clear insight on the mechanisms governing a
physical phenomena. In this work, we describe a proposal to use a
sophisticated combination of traditional Machine Learning methods to
obtain an explainable model that outputs an explicit functional
formulation for the material property of interest. We demonstrate the
effectiveness of our methodology in deriving a new highly accurate
expression for the enthalpy of formation of solid solutions of
lanthanide orthophosphates.
\end{abstract}

\begin{keywords}
Explainable learning, enthalpy, LASSO, ridge regression, sparsification, solid solutions.  
\end{keywords}

\begin{MSCcodes}
68T05, 62J07, 
\end{MSCcodes}

\section{Introduction}
\label{sec:I}

In recent years, the role of existing Machine Learning (ML) methods
have experienced a tremendous growth in many scientific computing
domains including Materials Science and Quantum Chemistry
\cite{Mjolsness2001,Ramprasad2017,Sutton20}. Concurrently, new
revolutionary methods and algorithms have appeared that expanded the
range of applicability of existing state-of-the art techniques
\cite{LeCunYann1998,Bolme2015}. This trend led to a high interest in
general ML applications which, in turn, left the scientific community
struggling in reconciling the need of developing refined tools with
the assessment of their usefulness when applied to specific problems
\cite{Wagstaff2012}. The assessment of their efficacy is particularly relevant when the target is an {\it explainable} learning method \cite{Roscher20} and the domain knowledge is {\it integrated} in the final model \cite{vonRueden21}.

 In recent years several machine learning methods have been proposed to predict enthalpies of formation of several categories of materials \cite{Stefanovic12,Deml16,Ubaru17}. Despite their progress, none of these efforts provides a fully explainable model which outputs a mathematical expression building on existing knowledge and improves it by adding to it additional terms that are statistically inferred.

In this article we propose a 3-step approach to use traditional machine learning tools to arrive at a scientifically consistent explainable method \cite{Roscher20}. At first, we propose the use of {\it Kernel Ridge Regression} \cite{Hofmann08} methods to first assess which, out of a number of different kernels, provides the most reliable and transparent method. Second, we make a post-hoc interpretation of the model and we proceed to {\it reverse engineer} it so as to find which coefficients of the model are the most relevant to recover a fully explainable mathematical expression for the target property. Finally, we integrate domain-specific knowledge by forcing scientific consistency through a constraint on how the input variables could be combined. The end result is a mathematical expression which relates the target property to the input variables in a functional dependence that replicates known results and add further terms substantially improving the accuracy of the final expression.       

Our methodology is in part inspired by the work of Ghiringhelli et al. \cite{Ghiringhelli2015, Ghiringhelli17} which uses the \textit{Least Absolute Shrinkage and
  Selection Operator} (LASSO) \cite{Santosa1986} together with a sparsification process they term LASSO$+\ell_0$ to learn the most relevant descriptors for a given target property. In this work, we go beyond their approach by constraining the functional form of the prior using knowledge coming from both the algorithmic model (the assessment of best kernel) and integration of domain-knowledge (to ensure scientific consistency). To demonstrate the feasibility of our approach we applied it to the specific problem of computing the excess enthalpy of formation of solid solution (enthalpy of mixing) of lanthanide orthophosphates ($Ln\rm PO_4$). We investigate the functional dependence of the mixing enthalpy for binary solid solutions of two distinct lanthanide cations ($Ln$), taking into account two distinct phases these materials form: monazite and xenotime \cite{JKK19}. 

\subsection{Excess enthalpy of solid solution formation}
\label{sec:entpy}

Monazite is a phosphate mineral that contains rare-earth elements.
Among these, lanthanide phosphates ($Ln\rm PO_4$) are the most widely distributed.
These form monazite-type structure for $Ln=\rm La,...,Tb$ and xenotime-type (zircon) structure 
for heavier lanthanides \cite{CPD11,NHM95,NAJ17,SHN13}. 
Among other potential applications, synthetic (monazite-type) ceramic solid matrices are suitable for the conditioning of long-lived radionuclides such as minor actinides (Np, Am, Cm) or Pu in a form of a synrock \cite{EW02,W00,MG00}.
However, before these ceramics could be used in nuclear
waste management, their physical and chemical properties and, most
importantly, their thermodynamic parameters 
have to be well characterized and understood. 

A solid solution is formed when two or more cations are incorporated into a solid host matrix on the same crystallographic site.
When atoms of the solute solid phase are incorporated into the solvent
solid host phase, the whole process can be interpreted as a sort of impurity inclusion into the host phase \cite{G00,PR09}. 
Here, we consider a combination of two cations within a single phase, either monazite or xenotime. In reality, however, when lighter (monazite as stable structure) and heavier (xenotime as stable structure) lanthanide are mixed, such a solid solution has a wide miscibility gap, i.e. it is thermodynamically unstable in a wide solid solution range, with different stable phases of solid solution endmembers (solute and solvent). In these cases, the mixing enthalpy of single phases solid solutions is a key factor for describing the two-phase cases, such as monazite-xenotime system \cite{M07,JKK19}.

Whether a single phase solution will stay stable or not, the result is driven by the
excess enthalpy of the mixing \cite{G00,LKB14}. 
The latter is defined as the difference between the formation enthalpies of the mixed compounds and those of the solid solution endmembers, which could be measured \cite{UHN01} or accurately computed \cite{BKB14,R12,BBK16}.
The single phase solid solutions such as monazite-type, resemble closely a symmetric solid solutions and are well described by a simple model, $H^E = m(1-m)W$, with $W$ being the Margules interaction parameter and $m$ the solid solution ratio \cite{PKG06,LKB14}. 
With systematic DFT-based calculations, it was found \cite{KL16}
that for the monazite-type solid solutions, the Margules interaction parameter $W$ is a function of the Young's modulus, $Y$,
the unit-cell volume, $V$, and the unit-cell volume difference between the solid solution endmembers (solid and solute), $\Delta V$, namely \cite{JBL17}
\begin{equation}
\label{eq:E1}
W \sim\frac{0.154Y}{V}\Delta V^2\sim \frac{Y}{6V	}\Delta V^2 .
\end{equation}

The relationship between the excess formation of mixing and the physical parameters 
has been a topic of discussion of various studies \cite{M07,LKB14,KL16}. Among these, the ionic radius of the mixing cations is often used as the main discriminant parameter \cite{M07}. Such a choice, however, makes the thermodynamic parameters of the mixing only weakly material dependent. 
As such, the excess enthalpy of mixing of monazite- and xenotime-type solid solutions are described with very similar relationship as a function of $\Delta R/R$ (Fig. 6 of \cite{M07}). 

\begin{figure}[ht!]
     \centering
     \begin{subfigure}{0.45\textwidth}
         \centering
         \includegraphics[width=\textwidth]{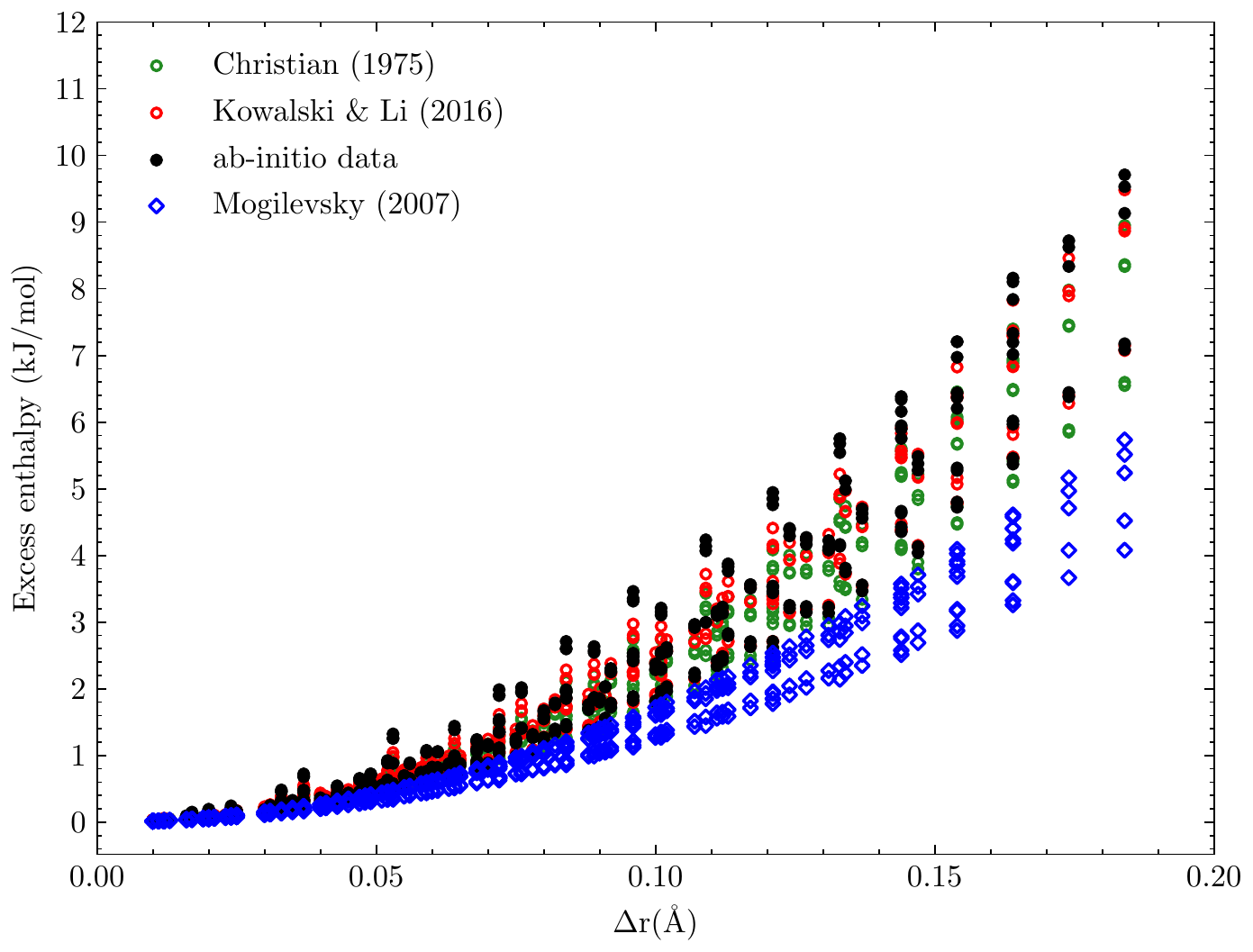}
        \caption{Monazite}
     \end{subfigure}
     \begin{subfigure}{0.45\textwidth}
         \centering
         \includegraphics[width=\textwidth]{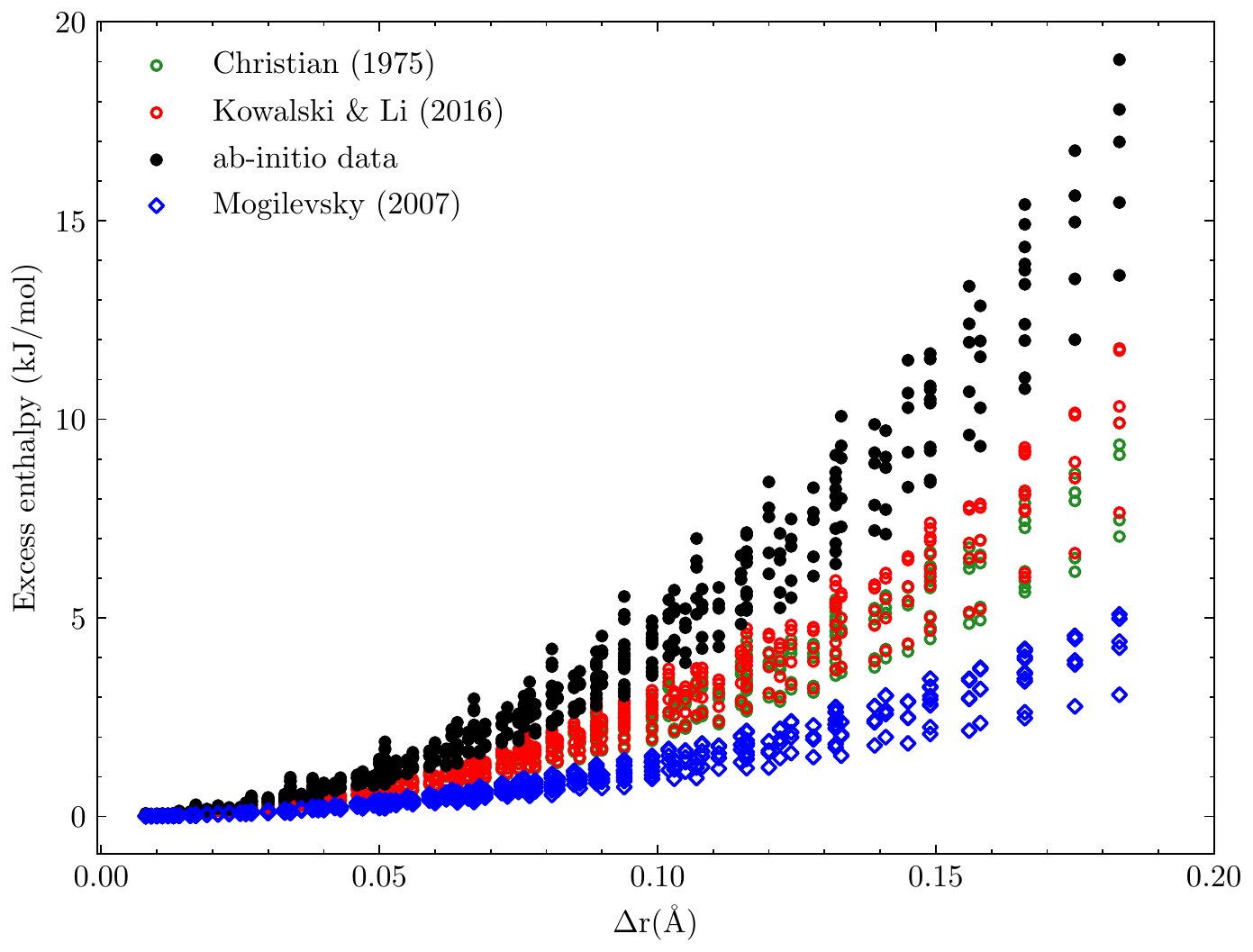}
         \caption{Xenotime}
     \end{subfigure}     
     \caption{The excess enthalpy of mixing for monazite- and xenotime-type solid solution computed {\it ab initio} and from models of \cite{C75}, \cite{KL16} and \cite{M07}}
\label{mon-xen}
\end{figure}

In Figure \ref{mon-xen} we illustrate how existing models describe the functional dependence on physical parameters of the excess of enthalpy  for the data used in this work. Plot (a) shows the case of monazite for which the models of \cite{KL16,C75} reproduce the data reasonably well. This is in part because both models use the difference in volumes of the endmembers as a parametric variable, while the model of \cite{M07} uses the difference in ionic radii. 

The situation is diametrically different in the case of xenotime-type solid solutions (plot (b) of Fig.~\ref{mon-xen}). Here, the models of \cite{C75} and \cite{KL16} give predictions that are inconsistent with the {\it ab initio} data by a factor of $\sim 2$. This points towards the possibility of another, unaccounted term in the Margules parameter that could be quite relevant in the case of xenotime-type solid solutions but of minor importance for monazite-type solid solution. 

A combination of {\it ab initio} and calorimetric studies, \cite{NKA17}
has shown that the {\it ab initio} data themselves are not enough to
constrain the values of the Margules parameter $W$, and that
understanding of the dependence of $W$ on the selected physical parameters
is crucial for precise modeling of the stability of 
solid solutions. 
As such, the study of the excess of enthalpy for this type of solid solutions lends itself perfectly to test the explainable machine learning methodology we have devised. 

\section{Methodology and learning algorithms}

In predicting materials' properties one needs a set of curated training data organized in {\it input} variables $\vec{x}$ and {\it target} properties $y$. 
The set of input variables has to represented in terms of a number of {\it independent} descriptors that are invariant under known mathematical and physical symmetries and usually requires a considerable amount of domain expertise. 
In this work, the input variables are represented by
properties of the elements constituting a solid solution (e.g. electron orbitals,
nuclear charge, etc.) and the target property by the solution excess enthalpy of formation.

\subsection{Elemental properties and descriptors}\label{Elemental properties and descriptors}

As mentioned in other works \cite{Ghiringhelli2015,
  Ubaru17,Stefanovic12} and based on the physics of the solution
process, properties associated to the electron orbitals and the nuclei
are expected to carry most of the weight in determining the value of
the enthalpy of formation. Moreover, all solid solutions that are part
of our data set have in common the same phosphate group
(PO$_4$). Consequently, properties of the atomic elements of such
group are not taken into consideration. Based on the online database
\url{http://www.knowledgedoor.com/}, we list in
Table \ref{tab:features} a meaningful list of elemental properties
({\it elementals}) $\epsilon_k$
that are available for each and every lanthanide element. 

\begin{table}[ht]
\centering
\begin{tabular}{c|cc}
\toprule
Name&Symbol&Unit\\ \midrule
Atomic number&$Z$& - \\
Atomic mass&$m$&[u]\\
Ionic radius coordination 8, 9&$R^8$, $R^9$&[\si{\angstrom}]\\
Ionization potential +2, +3&$IP^{2+}$, $IP^{3+}$&[eV]\\
Pauli electronegativity&$\chi$& - \\
Young's modulus of LnPO$_4$&Y&[GPa]\\
Effective nuclear charge&$Z_{eff}$& - \\
Macroscopic density&$\rho$&[g/cm$^3$]\\
Volume of LnPO$_4$& $V$ & [\si{\angstrom}$^3$] \\
\bottomrule
\end{tabular}
\caption{List of elemental properties and their physical units}
\label{tab:elementals}
\end{table}

These elementals can be arranged in an abstract vector
$\vec{\epsilon} = (\epsilon_1, \epsilon_2, \dots)\tran = (Z, m, R^8, R^9, IP^{2+}, IP^{3+}, \dots)\tran$. For each
lanthanide $L_i \in (\text{La}, \text{Ce}, \text{Pr} \dots)$
there is one such vector. 
We build descriptors out of elementals. Since we are investigating
solid solutions made of two lanthanides, our descriptors
$x_k[\epsilon_k(L_i), \epsilon_k(L_j), m_i,m_j]$ are functions of
elementals from two different lanthanides together with their mixing
ratio $m_i$. The inclusion of $m_i$s is necessary to distinguish
between different solution ratios.
However the descriptor is defined, it should be
invariant to simultaneous permutation of lanthanide and mixture
ratios. Noting that $m_i + m_j =1$, we can actually use only one
mixing ratio $m= m_i$ so that $m_j=1-m$.

Invariance under permutation can be expressed as
$$x_k[\epsilon_k(L_i), \epsilon_k(L_j), m] = x_k[\epsilon_k(L_j),
\epsilon_k(L_i), 1-m]$$
It is important that a descriptor $x$ changes
significantly with the mixing ratio. Additionally, we should
include descriptors capturing certain processes were the enthalpy is
strongly dependent on which lanthanide has the largest abundance. Last
but not least, descriptors has to be homogeneous functions of
elementals and cannot mix elementals with different physical units (unless
conversion constants are involved). For the reasons above we selected
three type of descriptors $\des{1}$, $\des{2}$ and $\des{3}$, listed
in Table \ref{tab:features}, for every elemental
$\epsilon_k$ and every lanthanide pair $(L_i , L_j)$.
\begin{table}[ht]
\centering
\caption{\label{tab:features}Descriptor types depending on lanthanide pairs ($L_i, L_j$), elemental $\epsilon_k$, and mixing ratio $m$.}    
\begin{tabular}{c|c}
\toprule
Name&Descriptor\\ \midrule
Weighted mean& $\des{1}_k = m\epsilon_k(L_i) + (1-m)\epsilon_k(L_j)$\\
Quadratically weighted mean&$\des{2}_k = \frac{m^2 \epsilon_k(L_i) + (1-m)^2 \epsilon_k(L_j)}{m^2 + (1-m)^2}$\\
Absolute difference&$\des{3}_k = |\epsilon_k(L_i) - \epsilon_k(L_j)|$\\
\bottomrule
\end{tabular}
\end{table}

Notice that the quadratically weighted mean is not quadratic in the
actual values of the elementals $\epsilon_k$ but quadratic in the mixture ratio
$m$. That means the descriptor $\des{2}$ will lean heavily to the value of
the elementals of the more abundant lanthanide. For each combination of
lanthanides pairs the ten elementals $\epsilon_k$ are organized in a descriptor
vector $\vec{x}$ made of thirty descriptors in total.
\begin{align*}
\vec{x}(L_i, L_j, m) = \begin{bmatrix}
\des{1}_1[\epsilon_1(L_i), \epsilon_1(L_j), m] \\
\des{2}_1[\epsilon_1(L_i), \epsilon_1(L_j), m] \\
\des{3}_1[\epsilon_1(L_i), \epsilon_1(L_j), m] \\
\des{1}_2[\epsilon_2(L_i), \epsilon_2(L_j), m] \\
\vdots
\end{bmatrix}
\end{align*}
Each vector $\vec{x}\in\mathcal{X}$ of size $d=30$ makes up the
input variables for the learning algorithm. The target value $y$ is
the excess enthalpy of formation $H^E$. 
For each choice of lanthanide
pairs $(L_i,L_j)$ and choice of mixing ratio $m$ we have a data
point $(\vec{x}, y)$. All data points together constitute
a set holding $N$ data points. We will see at the
end of this section how this set of points is generated. 


\subsection{Learning algorithms}
\label{sec:algo}
Since the data points in our set have both input values and
target value, we use a common type of {\it supervised} learning
algorithm: {\it kernel ridge regression} (KRR) \cite{Hofmann08}. Kernel ridge regression is a non-linear regression algorithm with a regularization
term (from which the name ``ridge'') that is comparable to the
well-known Support Vector Machine algorithm.

The simple linear regression algorithm aims at finding the unknown
coefficients $\vec{\beta}$ of the function
$f(\vec{z}) = \langle\vec{\beta},\vec{z}\rangle$ minimizing the error
$E[f(\vec{z}) - y]$ (also known as {\it loss function}) over the
entire set of data. In order to alleviate over fitting of the data, a
regularization term is usually added. In the {\it ridge regression},
the regularization amounts to adding a penalty term to the
minimization problem. Choosing the squared error as loss function
leads to the following minimization problem
\begin{equation}
\label{eq:ridge_regression}
\arg\min_{\vec{\beta}\in \mathbb{R}^d} \sum_i
  (\langle\vec{\beta},\vec{x}_i\rangle - y_i)^2 + \lambda\|\vec{\beta}\|_2^2  
\end{equation}

By introducing a function $\vec{z} \rightarrow \phi(\vec{z})$ which
maps the input space $\mathcal{X}$ to a feature space $\mathcal{H}$,
the use of kernels generalizes the linear to non-linear regression
(see \cite{Rupp15} for a didactic introduction). In this context a
kernel is an inner product in feature space
$k(\vec{x}_i,\vec{x}_j) = \langle\phi(\vec{x}_i), \phi(\vec{x}_j)\rangle$. The advantage of
using kernels is that the function
$f(\vec{z}) = \sum_i \alpha_i k(\vec{x}_i,\vec{z})$ is not expressed
anymore as a sum over dimension of the input space $d$, but instead as
a sum over the number of data $N$ making up the training set.  With
this set up, the minimization problem to be solved becomes
\begin{align*}
\arg\min_{\vec{\alpha}\in \mathbb{R}^N} \sum_i^N
  (f(\vec{x}_i) - y_i)^2 + \lambda\|f\|_{\mathcal{H}}^2.  
\end{align*}
In practice, the kernel function is expressed as a matrix of inner
products between points of the training data set in feature space
$k(\vec{x}_i,\vec{x}_j) = K_{ij}$. Eventually, the solution of
the minimization problem can be expressed by the linear equation
\begin{align}
\label{eq:alfa-coeff}
\vec{\alpha} = \left(\mat{K} +  \lambda \mat{1}_N\right)\inv\vec{y},
\end{align}
with $\vec{\alpha}\in\realsd{N}$ being the vector that contains the
information learned.

Almost all Machine Learning methods do not work directly out of the
box but have a number of parameters that have to be fixed. In
the case of the KRR, the level of regularization through the parameter
$\lambda$ 
is tuned for the
dataset at hand and the selected kernel. Additionally, almost every
kernel has some extra parameters that must also be tuned. The entire
set of these adjustable parameters are called \textit{hyperparameters}.

Given a kernel, its computation can still be performed in input space
despite its value describes the inner product in feature space. 
In this work, we employed three different kernels with the same set of
data: the polynomial, the Gaussian, and
the Laplacian kernels, which are respectively based on the inner
product, the $\ell_2$-norm and the $\ell_1$-norm
\begin{align}
k(\vec{x}, \vec{z}) = & \left(\langle\vec{x},\vec{z}\rangle + c\right)^p \label{eq:std-pol-kernel} \\
k(\vec{x}, \vec{z}) = & \exp\left(-\frac{\norm{\vec{x}-\vec{z}}^2}{2\sigma^2}\right)\\
k(\vec{x}, \vec{z}) = & \exp\left(-\frac{\norm[1]{\vec{x}-\vec{z}}}{\sigma}\right).
\end{align}
The actual computation of $\vec{\alpha}$ amounts to solve a positive-definite numerical
linear system $A \vec{\alpha} = \vec{y}$. 
Once $\vec{\alpha}$ is computed, it is used to compute predictions for any
new data point with
$\hat{y} = \sum_i \alpha_i k(\vec{x}_i,\vec{\hat{x}})$. For validation
purpose, the results of such prediction are typically presented in a
\textit{scatter plot} where the predicted and computed target values
are represented on the $x$ and $y$, respectively (see
plots of Fig.~\ref{Gauss3dpolyKernel}, for instance).

The Laplacian and Gaussian kernels are by far the most used in KRR because they provide the most effective map since they use a virtually limitless number of functions as the prior of our statistical model.
The down side is that we cannot recover the
explicit functional form which express the target value in terms of
input variables. This is why we have also included polynomial kernels, since they could be virtually inverted and return a functional expression for the coefficients $\vec{\alpha}$ in terms of descriptors.
Being able to statistically infer such functional
expression would allow us to go beyond the prediction of target values for new solid solutions and understand which descriptors are more relevant and contribute the most to determine the target values. 

In this sense, the polynomial kernels that will be closest in error to the Gaussian or Laplacian kernels will provide a hint on the order of polynomial functions that should be included in our prior. With this information we can manually construct thousands of candidate functions of the original descriptors $\vec{x_i^{(t)}}$ that could faithfully represent the underlying surrogate space.
In our application, we denote these candidate functions as $\vec{v} = [\vec{v_1}; \vec{v_2}; \cdots; \vec{v_M}]$, where $M$ can range
from $\mathcal{O}(10^3)$ to $\mathcal{O}(10^4)$.
We then apply a \textit{sparsification} technique, which amounts to find the most relevant among $\vec{v_k}$ by forcing as many coefficients of the statistical model to be zeros. In Section \ref{KRR Results}, justified by the KRR results, we show how each distinct $\vec{v_k}$ 
is constructed from a low degree $p$ polynomial function of the original descriptor. 

The objective of sparsification, is finding the most relevant term(s) among $\vec{v_k}$ which contributes the most to the target values. Moreover, the number of the relevant terms should be also controllable. A straightforward sparsification technique that one can employ is 
the LASSO \cite{Santosa1986} approach combined 
with an $\ell_0$ regularization. 
This combination is able to sparsify the coefficients of LASSO regression into a small determined number. The minimization problem to be solved 
is given as follows 
\begin{equation}
\label{eq:lasso_l0}
\arg\min_{\vec{\gamma}\in \mathbb{R}^M} \sum_i^N
  (\langle\vec{\gamma},\vec{v}_i\rangle - y_i)^2 + \hat{\lambda}\|\vec{\gamma}\|_0.  
\end{equation}

In this formula, the $\ell_0$-norm of a coefficient vector $\vec{\gamma}$ is 
defined as 
\begin{equation}
    \|\vec{\gamma}\|_0 = \#\{j: \gamma_j \neq 0\}, \quad j = 1, \dots, M
\end{equation}
indicating the number of non-zero elements of the $\vec{\gamma}$ vector.
A vector $\vec{\gamma} \in \mathbb{R}^M$, it is called $\kappa$-sparse 
if $\|\vec{\gamma}\|_0 \leq \kappa$.

While the LASSO$+\ell_0$ is the exact problem we want to solve, it has a significant drawback:
this minimization problem is not convex. This leads to a "NP-hard problem" \footnote{A NP-hard problem
means that the solution can not be found in polynomial time with respect to $M$.} which is infeasible 
when $M$ is large. Therefore,
LASSO$+\ell_0$ cannot be directly applied to sparsify the candidate functions $\vec{v}$. 
In order to compromise between the convexity and sparsity of the coefficient vector, we first utilize a LASSO with a Manhattan $\ell_1$-norm regularization $\lambda \|\vec{\gamma}\|_1$ (LASSO$+\ell_1$ in this paper) to carry out an initial \textit{feature selection} out of which we can achieve the sparsification of $\vec{v}$ \cite{Ghiringhelli17}.

\begin{equation}
\label{eq:lasso_l1}
\arg\min_{\vec{\gamma}\in \mathbb{R}^M} \sum_i^N
  (\langle\vec{\gamma},\vec{v}_i\rangle - y_i)^2 + \hat{\lambda}\|\vec{\gamma}\|_1.  
\end{equation}

This latter optimization problem is convex, which promote also sparsity. 
In equation (\ref{eq:lasso_l1}), $\hat{\lambda} > 0$ is the penalty parameter which is
able to control the sparsity of the coefficient vector $\vec{v}$: the larger $\hat{\lambda}$ is,
the smaller the $\ell_1$-norm $\|\vec{\gamma}\|_1$ would be, 
hence higher sparsity is achieved and more candidate functions are eliminated. 

In our application, we start from a very small $\hat{\lambda}$, then increase
it with a fixed step. With the increase of $\hat{\lambda}$, we observed that a large number
of constructed functions in $\vec{v}$ get quickly eliminated
to rapidly flatten the curve which stagnates even for much larger $\lambda$ values; in other words the minimization process with the $\ell_1$-norm reduces the vector $\vec{\gamma}$ to be at most   
$\kappa$-sparse, $\kappa$ being usually smaller than $20$ in our application. 
In practice, we compressed a very large feature space spanned by $\vec{v}$ 
into a tiny one spanned by $\vec{v'}$, in which $\vec{v'}$ is a subset of $\vec{v}$ whose
number of elements is smaller than $\kappa$. 
Thanks to the low dimension of feature space selected by LASSO$+\ell_1$, the Equation (\ref{eq:lasso_l0})
of LASSO$+\ell_0$ can be solved rather effectively with a brute force approach. 
More details of numerical results are given in Section \ref{results}.

This method, combining LASSO with a sparsification 
of the coefficients vector, has been developed
in the context of \textit{compressed sensing} \cite{Ghiringhelli17}, where one is interested
in reproducing the gathered data with as few as possible degrees of
freedom. Unlike in the \cite{Ghiringhelli17}, our work starts with a Kernel Ridge Regression 
and maps the descriptors $\vec{x}_i$ to a much larger but finite dimensional input space 
where the new features $\vec{v}_i$ are made out of a finite number of
selected functions of the original descriptors $\vec{x}_i$. 
We additionally integrate domain-knowledge by modifying the initial selection of descriptors and features to ensure scientific consistency with the existing knowledge from the physics of solid solutions. The necessity of tuning of the descriptors and the features is also confirmed by the inefficiency of the LASSO$+\ell_1$ method when the original descriptors $\vec{x}_i$ are used. 

In the next section, we see that KRR
polynomial kernel with degree $3$ returns one of the lowest statistical
error. We capitalize on such a result and make the plausible
hypothesis that a polynomial map of degree at most 3 for LASSO should
identify the most promising functions of the descriptors $\des{t}_k$. 
We then show that the results of \textit{feature selection} by LASSO$+\ell_1$ suggests the selection of a modified set of descriptors that guarantees scientific consistency. Finally we determine the desired functions which map the elemental descriptors to the excess of enthalpy for both monazite and xenotime.

\subsection{Data set generation}

The data set used by the selected learning methods was computed with the
Quantum-ESPRESSO code using the approach of \cite{LKB14,KL16}.  The
solid solutions were modeled with special quasi-random structures
constructed using procedure of \cite{ZWF90}. All the structures were
modeled with supercells containing 192 atoms (32 formula units).  We
applied the PBEsol exchange-correlation functional \cite{PRC08} and the
$f$ electrons were included into pseudopotential core. It has been
shown that this setup results in a very good predictions of structures and
thermodynamic parameters of lanthanide phosphates, including formation
enthalpies \cite{BKB14,JBB17,KBV15,BBK16}. The excess enthalpies of
mixing and Margules interaction parameters were derived from
differences in the total energies of the mixed cation structure and
weighted (according to the solid solution composition) sum of the end
members.

The dataset consists of excess enthalpies of formation between all 15
lanthanides, which leads to 105 possible combinations
($\sum_{i=1}^{14} i = 105$). Those 105 combinations were then modeled
for five different mixture ratios $m=0.25,\ 0.375,\ 0.50,\ 0.625,\ 0.75$
giving a total of $525$ data points. Two distinct data sets were generated
for the two lanthanide orthophosphate phases, monazite and xenotime, which correspond to the two possible coordination numbers of the lanthanides (see Table \ref{tab:elementals}). In the following, we will test our models on three distinct configurations of these data: monazite only ($525$ data points), xenotime only ($525$ data points), and fused ($1050$ data points). Not all these points are used to
train the learning model. A subset is reserved for testing and
validation purposes.


\section{Regression, sparsification and interpretable formulas}\label{results}

The first step before determining the coefficients $\vec{\alpha}$ of our KRR models is to
determine the optimal value of the corresponding hyperparameters. To this aim, the entire
given dataset is typically split in two parts. The \textit{training}
dataset $(\vec{x}_i, y_i)\in\mathcal{T}$ is used to compute the actual
coefficients $\vec{\alpha}$ of the regression that appear in
Eq.~\eqref{eq:alfa-coeff}.  The \textit{testing} or
\textit{prediction} dataset $(\vec{x'}_i, y'_i)\in\mathcal{P}$ is kept
separate and is used to evaluate the quality of a given set of hyperparameters
values for predicting unseen data
$\vec{\hat{y}} = \vec{\alpha}\tran\mat{K}'$, where $K'_{ij} = k(\vec{x}_i,\vec{x'}_j)$.
The optimal set of parameters is selected when the predictions
(i.e. $\hat{y}_i$) are in the best possible agreement with the known
values from the testing dataset (i.e. $y'_i$). Different error
functions are commonly used to quantifying the quality of a
prediction. The most common that we are also using in this work are 
Mean Squared Error (MSE), Mean Absolute Error (MAE), and Maximum Error (ME).
%
%

Because possible hyperparameters values may span over
multiple dozens of orders of magnitude, 
brute force methods that quickly scan the entire space in a grid-like pattern
are preferred to conventional minimization methods. Once an approximate local
minimum is found, local optimizations are used to refine the values
of the hyperparameters. In the following we do not report of the hyperparameter search results which is a standard procedure in the use of ML algorithms. In all the scatter plots and tables it is implicitly intended that all hyperparameters have been opportunely optimized following the procedure just described.

\begin{figure*}[ht]
     \centering
     \begin{subfigure}{0.49\textwidth}
         \centering
         \includegraphics[width=\textwidth]{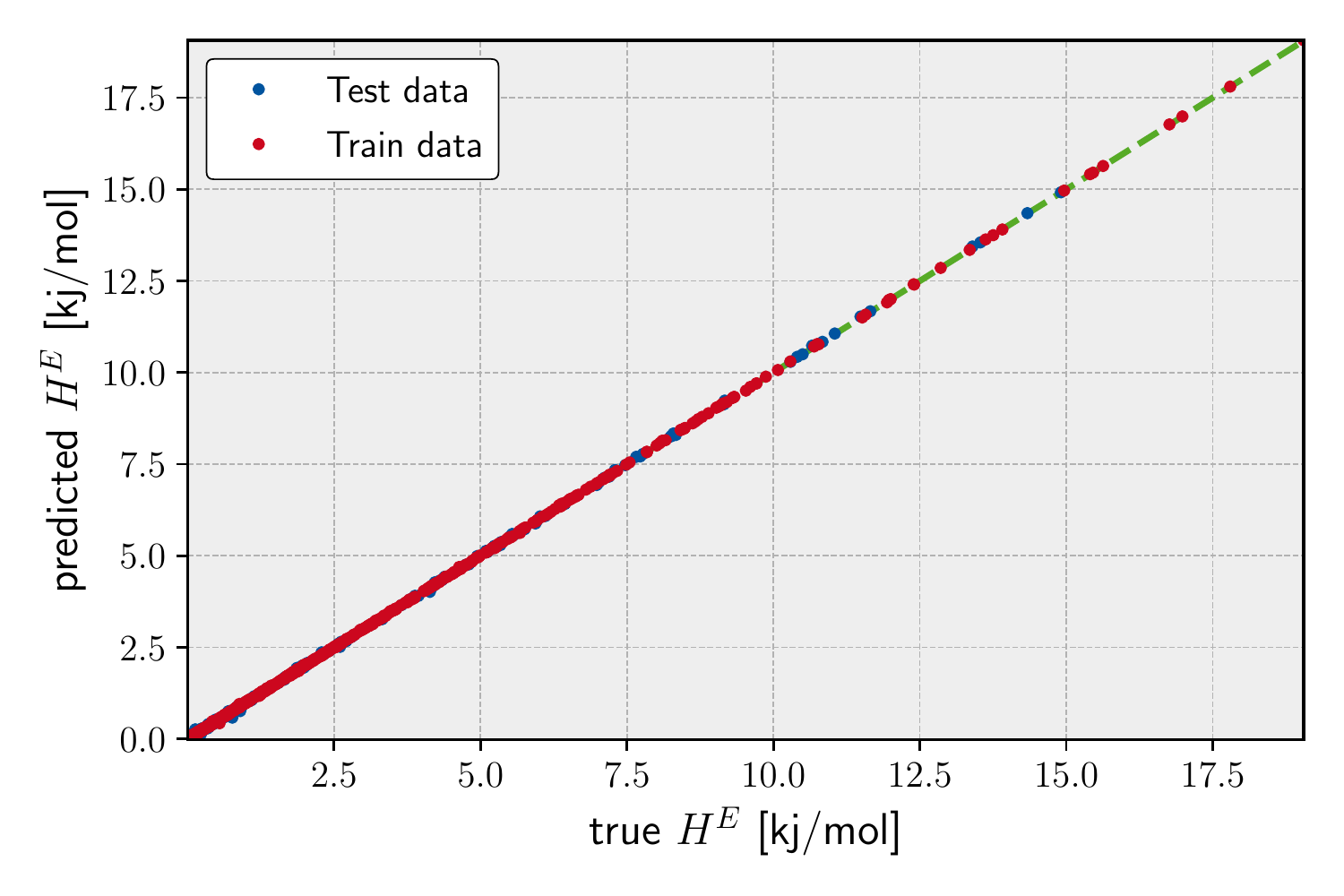}
     \end{subfigure}
     \hfill
     \begin{subfigure}{0.49\textwidth}
         \centering
         \includegraphics[width=\textwidth]{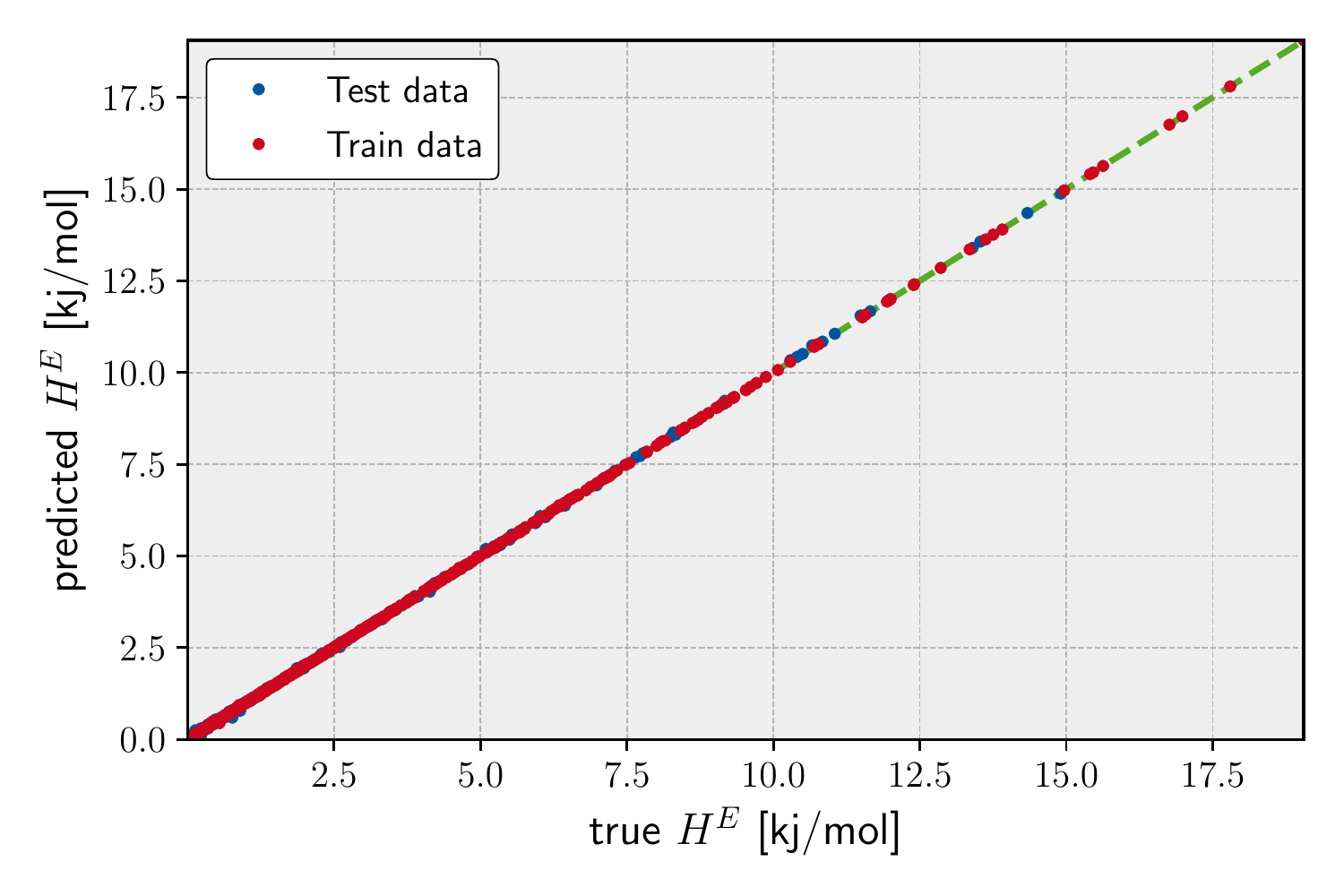}
     \end{subfigure}
     \caption{Regression for the Gaussian Kernel (left) and the
       degree 3 polynomial Kernel (right). The scatter plots
       corresponds to a trained model whose hyperparameters have been already fitted. Gaussian best fit: $\lambda =
       10^{-13.46}$, $\gamma = 10^{-5.90}$ with an MAE of
       $\SI{0.0107}{kJ/mol}$. Degree 3 polynomial the best
       fit: $\lambda = 10^{4.74}$,
       $\gamma = 10^{5.79}$, $c = 10^{0.37}$ with an MAE of
       $\SI{0.0090}{kJ/mol}$.}
\label{Gauss3dpolyKernel}
\end{figure*}

\subsection{Predicting excess of enthalpies with KRR}\label{KRR Results}
In the search for optimal hyperparameters, 
we have split the total subset of data in two parts between $\mathcal{T}$ and
$\mathcal{P}$. The ratio of the data size between $\mathcal{T}$ and
$\mathcal{P}$ is $4:1$.
%
Once the hyperparameters search is completed we fitted the data of the
training set $\mathcal {T}$ using all three distinct
kernels for the KRR model.

In Table \ref{RegressionResults} we report the results obtained for
all kernels with all three type of errors (MSA, MAE, ME) for both sets
$\mathcal{T}$ and $\mathcal{P}$ on the \textit{fused} data set configuration (similar results are obtained for the other two configurations).
Despite its remarkable low errors on
the $\mathcal{T}$ set, the Laplacian kernel does not return a
satisfactory result on the set of prediction data $\mathcal{P}$. In
fact, no choice of hyperparameters returns a reliable
regression for unseen data: the search space minimization return a
value for $\lambda$ numerically indistinguishable from zero. This is a
typical sign that KRR with this kernel is overfitting the data and
returning an in-sample error much smaller of the out-of-sample
error. For this reason we discarded the learning model using the
Laplacian kernel.

\begin{table*}[t!]
\centering
\small{
\begin{tabular}{c|ccc||c|ccc}
\textbf{Kernel} & $\log \lambda$ & $\log \gamma$ & $\log c$ & & \textbf{MAE} & \textbf{MSE} & \textbf{ME} \\\hline
\multirow{2}{*}{Poly. 2D} & \multirow{2}{*}{\num{-3.68}} & \multirow{2}{*}{\num{3.95}} & \multirow{2}{*}{\num{-0.447}} & $\mathcal{T}$  & 0.0098 &  0.0002 & 0.0917 \\
& & & & $\mathcal{P}$ & 0.0278 & 0.0025  & 0.4714 \\[1ex]
\multirow{2}{*}{Poly. 3D} & \multirow{2}{*}{\num{4.73}} & \multirow{2}{*}{\num{5.79}} & \multirow{2}{*}{\num{0.368}} & $\mathcal{T}$ & 0.0090 &  0.0001  & 0.1041 \\
& & & & $\mathcal{P}$ & 0.0228 & 0.0011  & 0.1865 \\[1ex]
\multirow{2}{*}{Gaussian} & \multirow{2}{*}{\num{-13.4}} & \multirow{2}{*}{\num{-5.90}} & \multirow{2}{*}{\texttt{-}} & $\mathcal{T}$ & 0.0107 &  0.0002  & 0.1140 \\
& & & & $\mathcal{P}$ & 0.0235 & 0.0011  & 0.1774 \\[1ex]
\multirow{2}{*}{Laplacian} & \multirow{2}{*}{\num{-20.0}} & \multirow{2}{*}{\num{-2.31}} & \multirow{2}{*}{\texttt{-}} & $\mathcal{T}$ & \num{7.6e-14}  &  \num{9.2e-27}  & \num{3.1e-13} \\
& & & & $\mathcal{P}$ & 0.1102 & 0.0272  & 0.6428 \\
\end{tabular}
}
\caption{Kernel Ridge Regression results on the Excess Formation Enthalpy dataset for different Kernel types. The first two rows are polynomial kernels of degree 2 or 3 respectively. The errors displayed are \textbf{Mean Squared Error} (MSE), \textbf{Mean Absolute Error} (MAE) and \textbf{Maximum Error} (ME). $\mathcal{T}$=Training dataset, $\mathcal{P}$=Prediction dataset. Units are [$\si{kJ^2/mol^2}$] (MSE) or [$\si{kJ/mol}$] (MAE \& ME). Note that the Laplacian kernel has by far the lowest regularization strength ($\lambda = 10^{-20}$) which leads to a perfect fit on the training data but by also to the worst performance on the testing dataset. This is a strong sign of overfitting.}
\label{RegressionResults}
\end{table*}

The low order polynomial and the Gauss kernels return a much nicer
picture in terms of the errors. Already the degree 2 polynomial kernel is able to fit the
data quite well. Its MAE differs only $\approx\SI{0.01}{kJ/mol}$ from the
error of the degree 3 kernel and the Gaussian Kernel. Judging from the
fact that the degree 3 polynomial kernel returns a MAE as low as
$\approx\SI{0.02}{kJ/mol}$ for the $\mathcal{P}$ set indicates that the underlying function for the
excess of Enthalpy could be represented by functions of the
descriptors having up to cubic terms. Since the actual values for the
excess enthalpy of formation for the lanthanide orthophosphates span
a range going from of $\num{0.5}$ to $\SI{10}{kJ/mol}$, the relative errors of our
model represent the same level of uncertainty returned by the DFT
simulations. In other words, the prediction provided by the KRR model
with either a degree 3 polynomial or the Gauss kernel are
indistinguishable from the finite accuracy of the in silico simulation
used to generate the data used in both sets $\mathcal{T}$ and
$\mathcal{P}$ (see Fig.\ref{Gauss3dpolyKernel}). 

What distinguishes the Gaussian and the degree 3 polynomial kernels are the value of the hyperparameters: the degree 3 kernel requires a quite large value of $\lambda$ and $\gamma$ respect to the Gaussian. This is not necessarily a negative result but it points out that our choice of descriptors may not be ideal when the kernel represents a finite set of prior functions like in the case of the degree 3 polynomial. When the set of prior functions becomes virtually infinite (the Gaussian kernel can be seen as an infinite series of polynomials), the descriptor choice becomes unimportant. We will see in the following subsection how the choice of descriptors becomes significant when one would like to sparsify the vector of coefficients $\vec{\alpha}$ starting from a finite set of prior functions of the descriptors.

\begin{table}[t!]
\centering
\begin{tabular}{p{2.6cm} p{1cm}|ccc}
Regressions ID & Set type & \textbf{MAE} & \textbf{MSE} & \textbf{ME}\\
\toprule  
\multirow{2}{*}{Original} & $\mathcal{T}$ & 0.0090& 0.0001  & 0.1041 \\
& $\mathcal{P}$ & 0.0228 & 0.0011  & 0.1865 \\ \midrule
\multirow{2}{*}{1} & $\mathcal{T}$ & 0.0126 & 0.0004  & 0.1705 \\
& $\mathcal{P}$ & 0.0120 & 0.0004  & 0.1865 \\ \midrule 
\multirow{2}{*}{2} & $\mathcal{T}$ & 0.0130 & 0.0004  & 0.1864 \\
& $\mathcal{P}$ & 0.0108 & 0.0002  & 0.0832 \\ \midrule
\multirow{2}{*}{3} & $\mathcal{T}$ & 0.0123 & 0.0004  & 0.0171 \\
& $\mathcal{P}$ & 0.0128 & 0.0005  & 0.1865 \\ \midrule
\multirow{2}{*}{4} & $\mathcal{T}$ & 0.0127 & 0.0004  & 0.1865 \\
& $\mathcal{P}$ & 0.0126 & 0.0004  & 0.1108 \\ \midrule
\multirow{2}{*}{5} & $\mathcal{T}$ & 0.0120 & 0.0004  & 0.1865 \\
& $\mathcal{P}$ & 0.0139 & 0.0005  & 0.1705 \\ \bottomrule
\end{tabular}
\caption{Errors for cross-validation of polynomial
  kernel of degree 3 with five different combinations of data points
  evenly split between $\mathcal{T}$ and $\mathcal{P}$. For each new
  regression, we used the same hyperparameters $\lambda$, $\gamma$ and
  $c$ as in the original one.}
\label{CrossValidation}
\end{table}

To ensure that our KRR models provide a good fit for all data
independently on how they are split between $\mathcal{T}$ and
$\mathcal{P}$, we have used {\it cross validation}.
In practice, we run the KRR models fitting several different choices
of training and testing datasets always keeping the same choices of
values for the hyperparameters that were previously selected. If the
results with the original dataset were truly a product of chance, a
fit with an entirely new combination of data points should show a
different performance. Table \ref{CrossValidation} shows the results
for repeating the KRR with polynomial kernel of degree 3 for five
different subsets, some of which even showed slightly better
performance than the original regression. The MAE over the five
different sets ranges between $0.052-0.082\ \si{kJ/mol}$, confirming the
quality of the original regression.

\begin{figure}[ht]
     \centering
     \begin{subfigure}{0.9\textwidth}
         \centering
         \includegraphics[width=\textwidth]{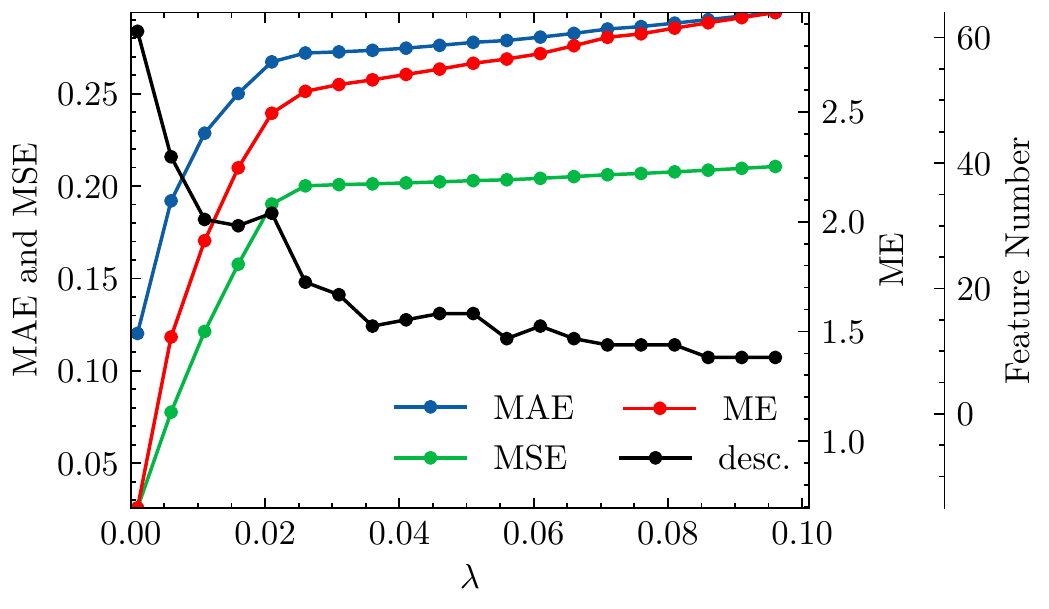}
         \caption{fused}
     \end{subfigure}
 
     \begin{subfigure}{0.49\textwidth}
         \centering
         \includegraphics[width=\textwidth]{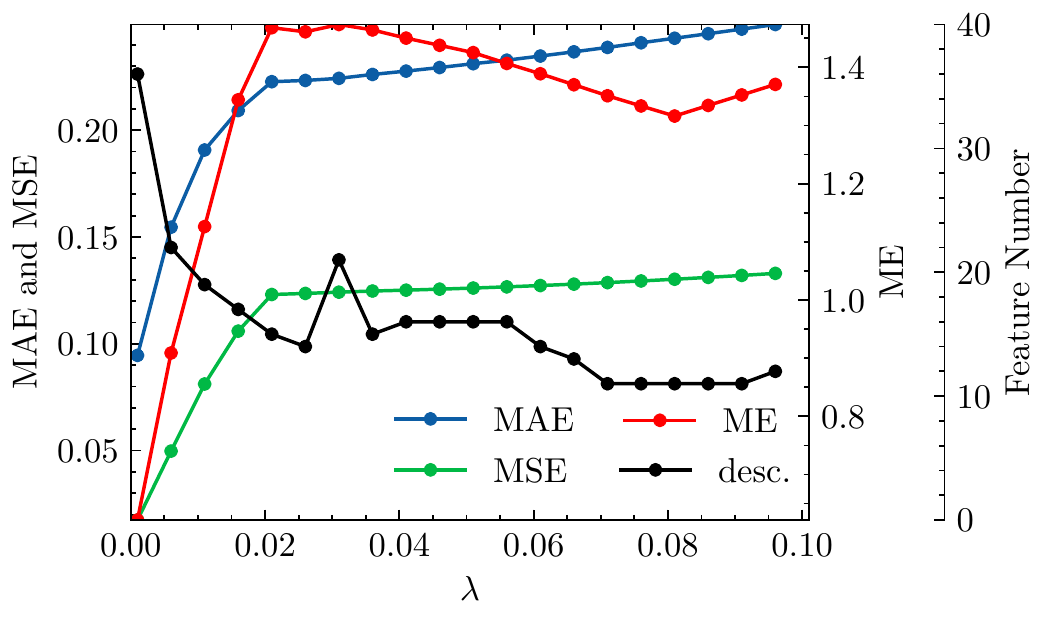}
        \caption{monazite}
     \end{subfigure}
     \hfill
     \begin{subfigure}{0.49\textwidth}
         \centering
         \includegraphics[width=\textwidth]{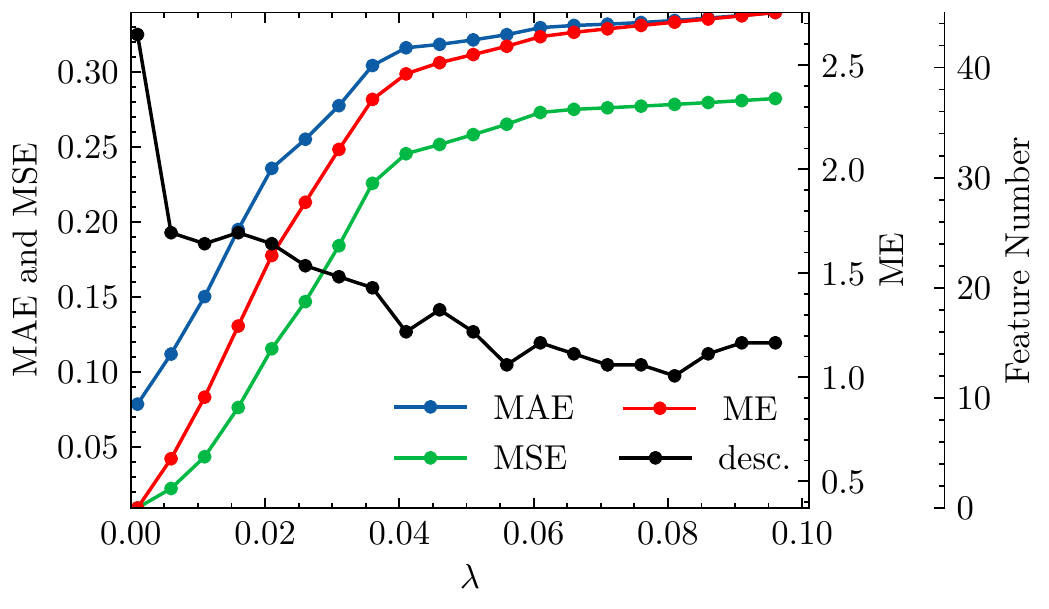}
                  \caption{xenotime}
     \end{subfigure}     
     \caption{LASSO$+\ell_1$ results with descriptors used in KRR}
\label{lassoKRRdescrs}
\end{figure}

\begin{figure}[ht]
     \centering
     \begin{subfigure}{0.9\textwidth}
         \centering
         \includegraphics[width=\textwidth]{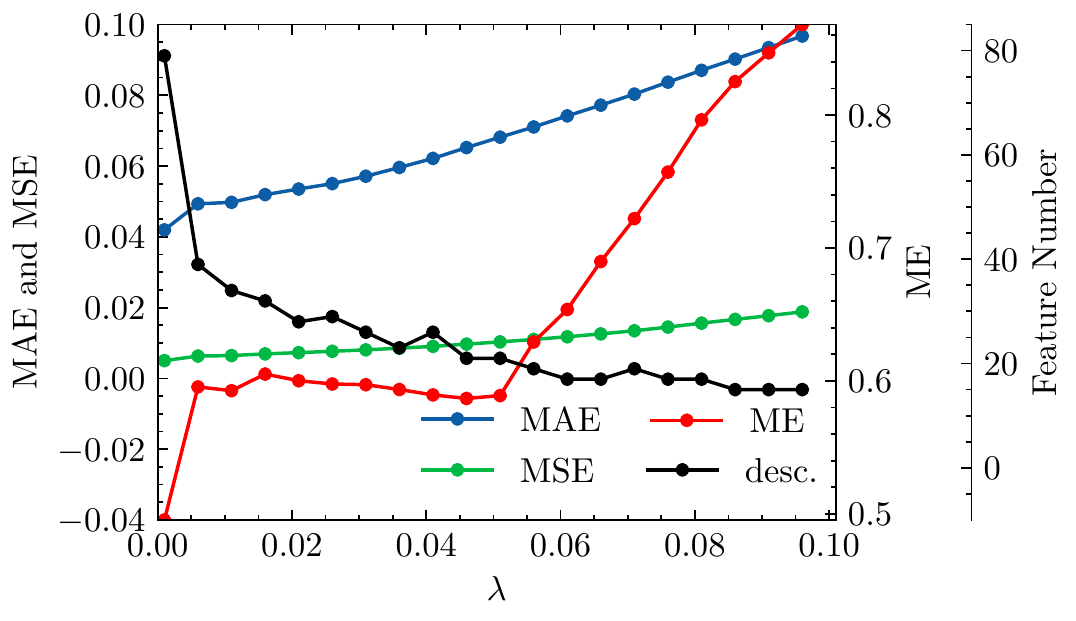}
                  \caption{fused}
     \end{subfigure}
     
     \begin{subfigure}{0.49\textwidth}
         \centering
         \includegraphics[width=\textwidth]{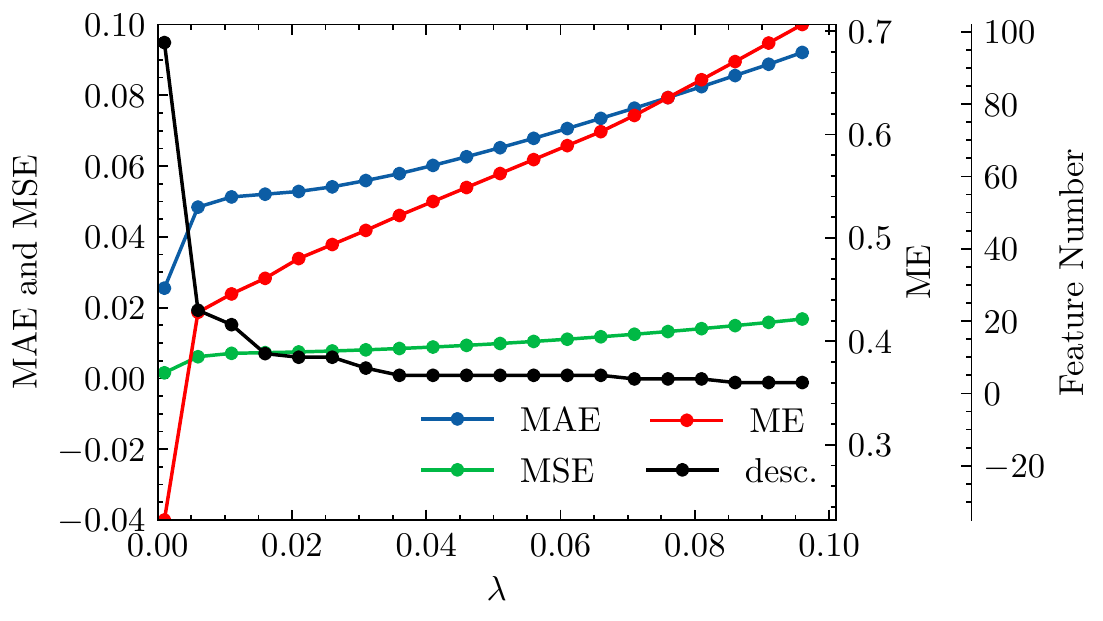}
         \caption{monazite}
     \end{subfigure}
     \hfill
     \begin{subfigure}{0.49\textwidth}
         \centering
         \includegraphics[width=\textwidth]{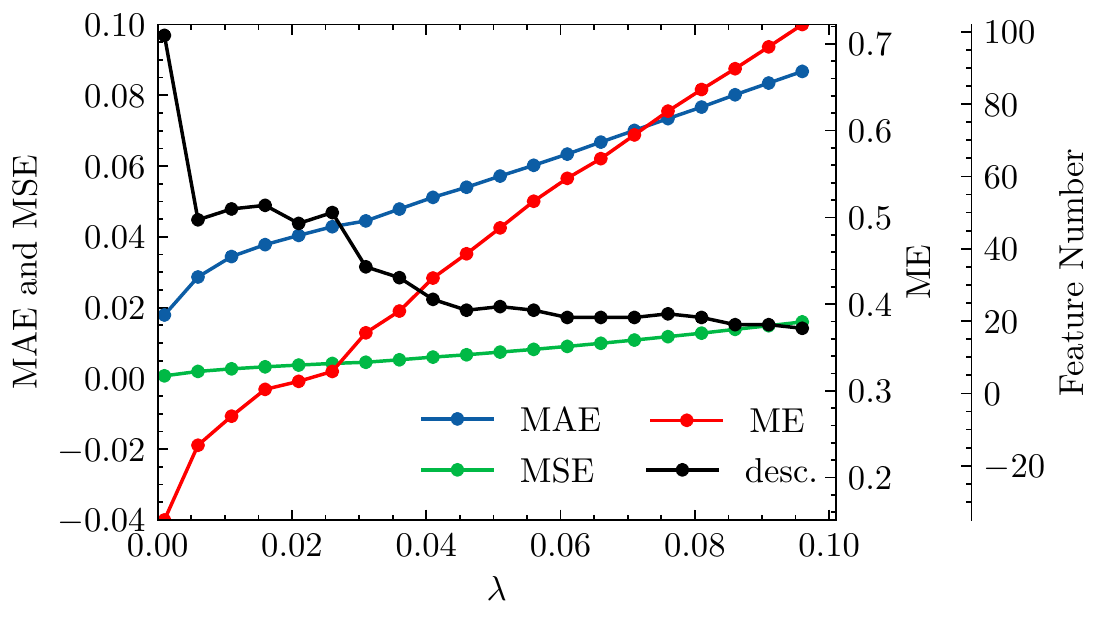}
         \caption{xenotime}
     \end{subfigure}     
     \caption{LASSO$+\ell_1$ results with descriptors defined only by arithmetic means and absolute difference.  }
\label{lassoNewdescs}
\end{figure}

\begin{table}[t!]
    \begin{subtable}[h]{\textwidth}
        \centering
\begin{tabular}{c|p{6.5cm}|ccc}
	nb. & Leading terms & \textbf{MAE} & \textbf{MSE} & \textbf{ME}\\
	\toprule  
	1 & $(x_{V}^{(3)})^2x_{V}^{(1)}$ & 0.3528 &0.4400 & 3.6657\\ \midrule
	2 & $(x_{V}^{(3)})^2$, $x_{V}^{(3)}x_{Y}^{(3)}x_{Y}^{(1)}$ & 0.2730 &0.2053 & 2.4429\\ \midrule
	3 & $(x_{V}^{(3)})^2$, $x_{V}^{(3)}x_{V}^{(2)}x_{Y}^{(2)}$, $x_{V}^{(3)}x_{Y}^{(3)}x_{Y}^{(1)}$ & 0.2757 & 0.2017 & 2.5846 \\ \bottomrule
	\end{tabular}
	\caption{fused}
	\end{subtable}

    \begin{subtable}[h]{\textwidth}
        \centering
\begin{tabular}{c|p{6.5cm}|ccc}
	nb. & Leading terms & \textbf{MAE} & \textbf{MSE} & \textbf{ME}\\
	\toprule  
	1 & $(x_{V}^{(3)})^2x_{R}^{(1)}$ & 0.2274 &0.1302 & 1.8870\\ \midrule
	2 & $(x_{V}^{(3)})^2x_{V}^{(1)}$, $(x_{V}^{(3)})^2x_{R}^{(1)}$ & 0.2201 &0.1237 & 1.4692\\ \midrule
	3 & $x_{Z_{eff}}^{(2)}$, $x_{R}^{(2)}x_{Z}^{(2)}$, $(x_{V}^{(3)})^2x_{\chi}^{(1)}$ & 0.2076 & 0.0799 & 1.1821 \\ \bottomrule
	\end{tabular}
	\caption{monazite}
	\end{subtable}
	
    \begin{subtable}[h]{\textwidth}
        \centering
\begin{tabular}{c|p{6.5cm}|ccc}
	nb. & Leading terms & \textbf{MAE} & \textbf{MSE} & \textbf{ME}\\
	\toprule  
	1 & $(x_{V}^{(3)})^2x_{Z}^{(1)}$ & 0.3303 &0.2865 & 2.3072\\ \midrule
	2 & $(x_{Z_{eff}}^{(3)})^2x_{\chi}^{(1)}$, $(x_{V}^{(3)})^2x_{Z}^{(1)}$ & 0.3247 &0.2784 & 2.4090\\ \midrule
	3 & $(x_{Y}^{(2)})^3$, $x_{R}^{(2)}x_{Z}^{(2)}$, $(x_{V}^{(3)})^2x_{Z}^{(1)}$ & 0.2078 & 0.0698 & 0.9729 \\ \bottomrule
	\end{tabular}
	\caption{xenotime}
	\end{subtable}
	\caption{LASSO$+\ell_0$ results for Fused, monazite and xenotime with descriptors used in KRR}
	\label{lassol0KRRdesc}
\end{table}

\begin{table}[t!]
    \begin{subtable}[h]{0.98\textwidth}
        \centering
\begin{tabular}{c|p{6.7cm}|ccc}
	nb. & Leading terms & \textbf{MAE} & \textbf{MSE} & \textbf{ME}\\
	\toprule  
	1 & $\frac{(\diff{V}\mean{V})^2}{\mean{R}^3}$  & 0.2798 &0.2268 & 3.6657\\ \midrule
	2 & $\frac{\diff{Y}\diff{Z}}{\mean{Z_{eff}}}$, $m^2(1-m)^2(\diff{V})^2$ & 0.0950 &0.0254 & 1.5519\\ \midrule
	3 & $\frac{\diff{Y}\diff{Z}}{\mean{Z_{eff}}}$, $m^2(1-m)^2(\diff{V})^2$, $(1-m)\diff{Y}(\diff{V})^2$ & 0.0878 & 0.0164 & 0.7348 \\ \bottomrule
	\end{tabular}
	\caption{fused}
	\end{subtable}

    \begin{subtable}[h]{0.98\textwidth}
        \centering
\begin{tabular}{c|p{6.7cm}|ccc}
	nb. & Leading terms & \textbf{MAE} & \textbf{MSE} & \textbf{ME}\\
	\toprule  
	1 & $m(1-m)(\diff{V})^2$ & 0.0622 &0.0095 & 0.4936\\ \midrule
	2 & $m(1-m)(\diff{V})^2$, $m^2(1-m)^2(\diff{V})^2$ & 0.0575 &0.0082 & 0.6309\\ \midrule
	3 & $m(1-m)(\diff{V})^2$, $m(1-m)^2(\diff{V})^2$, $(\frac{\diff{V}}{m})^3$ & 0.0516 & 0.0064 & 0.6195 \\ \bottomrule
	\end{tabular}
	\caption{monazite}
	\end{subtable}
	
    \begin{subtable}[h]{0.98\textwidth}
        \centering
\begin{tabular}{c|p{6.7cm}|ccc}
	nb. & Leading terms & \textbf{MAE} & \textbf{MSE} & \textbf{ME}\\
	\toprule  
	1 & $m(1-m)(\diff{V})^2$ & 0.1699 &0.0688 & 1.1805\\ \midrule
	2 & $m(1-m)(\diff{V})^2$, $\frac{(\diff{V})^2}{(\mean{V})^4}$ & 0.0793 &0.0237 & 0.9359\\ \midrule
	3 & $m(1-m)(\diff{V})^2$, $\mean{Z_{eff}}(\diff{V})^2$, $\frac{(\diff{V})^2}{(\mean{V})^4}$ & 0.0483 & 0.0051 & 0.4407 \\ \bottomrule
	\end{tabular}
	\caption{xenotime}
	\end{subtable}
	\caption{LASSO$+\ell_0$ results for Fused, monazite and xenotime with descriptors defined only by arithmetic means and absolute difference}
	\label{lassol0newdesc}
\end{table}

\subsection{LASSO + \texorpdfstring{$\ell_0$}{ell zero}}

In the previous section, we have concluded that KRR with Gaussian and degree 3 polynomial kernel performs very similarly with a slight advantage for the latter kernel. In this section we want to pursue the road of finding a surrogate model that is explainable: we aim at formulation of the excess enthalpy of mixing that can provide the domain scientist with a understandable function of a small number of descriptors. We are driven to recover this result by the observation that already polynomial functions of degree 3 provide enough prior to get an accurate KRR model. We achieve this result through a so-called {\it sparsification} process.

\subsubsection{Sparsification with Descriptors in KRR}\label{sparseKRR}

A large number of candidate functions
have been built from the polynomial kernel of degree $3$ based on the $27$ elemental descriptors introduced in Section \ref{Elemental properties and descriptors}. 
Denoting the group of $27$ elemental descriptors as $\mathcal{D}_1$, a group of candidate
functions with polynomial of degree $2$, denoted as $\mathcal{D}_2$, can be defined as a 
commutative element-wise vector product $d_id_j$, with $d_i, d_j\in \mathcal{D}_1$.  
The number of descriptors in $\mathcal{D}_2$ is $378$. The group of candidate functions
with polynomial of degree $3$, denoted as $\mathcal{D}_3$, can be defined in a similar
way, as a commutative element-wise product of three vectors $d_id_jd_k$ with $d_i$, $d_j$, $d_k\in \mathcal{D}_1$. The number of descriptors in $\mathcal{D}_3$ is \num{3303}. Therefore, all the 
candidate functions are a union of $\mathcal{D}_1$, $\mathcal{D}_2$ and $\mathcal{D}_3$, which is
denoted as $\mathcal{D}$. $\mathcal{D}$ is a dense matrix of size $\num{1050} \times \num{3708}$ for the fused data set, and of size $525 \times \num{3708}$ for both monazite and xenotime data set configurations. 

As described in Section \ref{sec:algo}, a \textit{feature selection} step has been performed on $\mathcal{D}$ through the LASSO$+\ell_1$ 
method. Increasing the penalty parameter $\lambda$, the size of feature space is reduced as more candidate functions are removed. In the numerical experiments, $\lambda$ is increased from \num{0.001} to \num{0.096} in incremental steps of \num{0.005}. We performed the same feature selection step
not only for the fused data, but also separately for monazite and xenotime data configurations. The 
results are plotted in Fig. \ref{lassoKRRdescrs}, which shows the changes of the errors (MAE, MSE, ME) and 
the size of the reduced feature space (marked as desc.).

From Figure Fig. \ref{lassoKRRdescrs} we infer that with the increase of $\lambda$, the feature space size can be
quickly reduced to less than $30$. This is good news, since sparsifying a feature space of size $\sim \num{30}$ through LASSO$+\ell_0$ into a determined size smaller than $5$ is still feasible \footnote{This is an NP-hard problem which can still be solved by brute-force for such small size}. At the same time, we observe 
that the all the errors increase with the increase of $\lambda$. This behavior is to be expected and is the direct consequence of the reduction of the feature space size. Unfortunately, the errors increase too quickly and show an exponential growth that plateaus already for small values of $\lambda$. For example, for \textit{fused} data, the MAE
increases from $\sim$ \num{0.03} to more than \num{0.25} when $\lambda$ is only increased from \numrange{0.001}{0.021}. All the while the MSE and ME increase to \num{0.20} and \num{2.5}, respectively. When increasing $\lambda$ from $0.021$
to $0.096$, the feature space size continues to be reduced, however, the changes of error flatten. Similar trends can also be observed separately for the monazite and xenotime data. This result signals that the generated feature space doesn't capture well the functional dependence of the excess enthalpy on the descriptors. In other words, we cannot find a simple functional dependence of $H^E$ and need many hundreds of functions to faithfully predict the enthalpy. 


In order to confirm our concerns, we perform a LASSO$+\ell_0$ step on the reduced feature space 
separately for the fused, monazite, and xenotime data set configurations. At most three leading
terms have been selected, which are listed in Table \ref{lassol0KRRdesc}, along with their corresponding
MAEs, MSEs, and MEs. As expected, the errors of candidate functions selected by LASSO$+\ell_0$
are more than one of order of magnitude larger than the ones derived by KRR. What is more remarkable is the fact that the errors don't seem to decrease much as we consider more terms. In fact, for the fused data configuration, the MAE and ME actually increase going from 2 to 3 leading terms. The errors for the other data configuration seems a bit better but are still far from what we would like to observe. From these tables and the plots in Figure \ref{lassoKRRdescrs}, we conclude that the choice of prior candidate functions cannot be effectively exploited by the sparsification process. 

\subsubsection{Sparsification with Descriptors built with Prior Knowledge}
\label{sec:sparse-knowledge}

In order to overcome the shortcomings of the sparsification process seemingly caused by the choice of descriptors, we simplify the form of the descriptors and exploit the existing knowledge from the application domain. Based on the insight provided by the simple model with the Margules interaction parameter $W$ and the expression in Eq. \eqref{eq:E1}, we make three additional hypothesis: i) the polynomial degree of $m$ and $1-m$ may not be in accord with each other and the polynomial degree of the elemental property $\epsilon_k$, ii) negative power of the descriptors may appear in the Margules parameter and, iii) the volume $V$, coordination number $R$ , and mixing ratio $m$ play a special role than the other elemental properties and may contribute with monomials of degree higher than 3. 

The direct consequence of i) is that $m$ and $1-m$ have been decoupled from the elemental descriptors and included as descriptors on their own. This may seem a strange choice since they do not depend on the lanthanides but are the same for all. On the other hand, decoupling the mixing ratio allows more freedom in the way it appears in the degree 3 polynomial functions that are part of $\mathcal{D}$. The indirect consequence is that we do not have anymore three types of descriptors 
for each elemental property $\epsilon_k$ 
 but only two: we drop the weighted quadratic mean, convert the weighted mean to a simple arithmetic mean, $\mean{\epsilon_k}$ and keep the absolute difference $\diff{\epsilon_k}$ (see Table \ref{diffAndmean}). 
 
 The second hypothesis, implies that the inverse of each elemental property $\diff{\epsilon_k}$
and $\mean{\epsilon_k}$ are also included explicitly in the descriptors.
Additionally, due to the third hypothesis, we include as descriptors monomials of degree higher than one for $m$, $V$, and $R$. In particular, for $m$ and $1-m$ we include up to quadratic terms (and their inverse), and for $V$ and $R$ we include up to cubic terms (and their inverse)  


\begin{table}[htbp]
\centering
\caption{Two basic descriptors of elemental property $\epsilon_k$}\label{diffAndmean}
\begin{tabular}{c|c}
\toprule
Name&Descriptor\\ \midrule
Absolute difference&$\diff{\epsilon_k} = \frac{|\epsilon_K(L_i) - \epsilon_K(L_j)|}{2}$\\
Arithmetic mean & $\mean{\epsilon_k} = \frac{|\epsilon_k(L_i) + \epsilon_k(L_j)|}{2}$\\
\bottomrule
\end{tabular}
\end{table}

All these descriptors build up a basic feature space $\mathcal{D}_1$ of size  $58$. Analogously as done for the original descriptors, the group $\mathcal{D}_2$ and $\mathcal{D}_3$ are built as the element-wise product of two or three features out of $\mathcal{D}_1$, respectively. Feature space $\mathcal{D}_2$ is of size $1647$ after removing $6$ features with standard deviation being $0$, such as the element-wise product of $\frac{1}{m}$ and $m$. The size of feature space $\mathcal{D}_3$ is $30856$. The sum of all candidate functions are collected into 
$\mathcal{D}=\mathcal{D}_1 \cup \mathcal{D}_2 \cup \mathcal{D}_3$. The size of data $\mathcal{D}$ 
is $1050\times32561$ and $525\times32561$ for fused and for monazite/xenotime data configurations.

We run the same LASSO$+\ell_1$ as in Section \ref{sparseKRR}, the results of which are shown in Fig. \ref{lassoNewdescs}. Compared to the Fig.~\ref{lassoKRRdescrs} the results we obtained are quite more promising. For fused data in Fig. \ref{lassoNewdescs}, the size
of reduced feature space drops down quickly below $40$ already at very small value of $\lambda$. Afterwards, with 
the increase of $\lambda$, the decrease of the size of reduced feature space slows down, which implies that the reduced feature space contains always important features for the target problem. 
Meanwhile, with the increase of $\lambda$, the errors increase moderately and linearly, which is a second sign that the feature space is reduced into a good choice of representative functions. 

After applying LASSO$+\ell_0$ on the vector of descriptors spanning the reduced space, the first three leading terms and their corresponding errors for fused, monazite and xenotime data configurations are shown in Table \ref{lassol0newdesc}. It is important to notice that while the errors may still be quite high for keeping only the leading term, they decrease rapidly when we include higher terms. Moreover the first leading term for the monazite and xenotime data configurations resemble very closely the expression of Equation \eqref{eq:E1}. We will analyze in more details the results and interpret their physical meaning up to five leading terms in the next section.


\section{Numerical results}

As illustrated in Sec.~\ref{sec:entpy}, the discrepancies between existing models and the data makes the computation of the excess of enthalpy for solid solutions of both monazite and xenotime, a clear cut example to demonstrate the validity of our statistical approach in retrieving an explainable expression for $H^E$. We will (1) show which descriptor provides the most reliable leading term for the mixing enthalpy between ionic radii of the mixing cations and the volumes of the pure phases, (2) identify the first sub-leading term, which enhances the accuracy of the $H^E$ prediction of xenotime-type solid solutions, (3) provide a exhausting statistical analysis for additional sub-leading terms.

\begin{table}[htbp]
\caption{Fused: all elemental descriptors except Ionic radius coordination}
\begin{tabular}{|c|l|} \hline
\# & Functions\\  \hline

1 & $\num{72.5497}\frac{\diff{V}\diff{Y}}{\mean{IP^{2+}}} + \num{0.3661}$\\  \hline
2 & $\num{302.2691}\frac{\diff{V}\diff{Y}}{\mean{Y}}+ \num{14068.2423}m^2(1-m)^2(\diff{V})^2 - \num{0.0271}$ \\ \hline
3 & $\num{267.3993} \frac{\diff{V}\diff{Y}}{\mean{Y}} + \num{13775.7831}m^2(1-m)^2(\diff{V})^2 + \num{7.1386}(1-m)\diff{Y}(\diff{V})^2 $ \\
& $ + \num{0.0135}$\\ \hline
4 & $\num{22.9437}\frac{\diff{V}\diff{Y}}{\mean{IP^{2+}}} + \num{13090.1088}m^2(1-m)^2(\diff{V})^2 + \num{5.9124}(1-m)\diff{Y}(\diff{V})^2 $ \\
& $  + \num{185.5363}\frac{(\diff{V})^2\diff{Y}}{\diff{M}} + \num{0.0040}$\\ \hline
5 & $\num{0.3198}m^2(1-m)^2\diff{Y} + \num{0.0001}\mean{IP^{3+}}\diff{IP^{3+}}\diff{Y} + \num{0.0335}\diff{V}\mean{IP^{3+}}\diff{Y} $\\                  
  & $+ \num{12977}m^2(1-m)^2(\diff{V})^2+ \num{7.3732}(1-m)\diff{Y}(\diff{V})^2 - \num{0.0603}$ \\ \hline
\end{tabular}

\vspace{10pt}

\begin{tabular}{|c|ccccc|} \hline
\diagbox{Error}{Desc. \#}&
\textrm{1}&
\textrm{2}&
\textrm{3}&
\textrm{4}&
\textrm{5}\\ \hline

MAE & 0.5581 & 0.1374 &  0.1249 &  0.1110 & 0.1075\\ \hline
MSE & 0.6803 & 0.0526 & 0.0403 & 0.0329 & 0.0260\\ \hline
ME & 4.0386 & 1.7908 & 1.5139 & 1.1327 & 1.2419\\ \hline
\end{tabular}
\label{S10}
\end{table}

\begin{table}[htbp]
\caption{Fused: all elemental descriptors except volume}
\begin{tabular}{|c|l|} \hline
\# & Functions\\ \hline

1 & $\num{72.5497}\frac{\diff{R}\diff{Y}}{\mean{IP^{2+}}} + \num{0.3661}$\\ \hline
2 & $\num{302.2691}\frac{\diff{R}\diff{Y}}{\mean{Y}} + \num{14068.2423}m^2(1-m)^2(\diff{R})^2 - \num{0.0271}$ \\ \hline
3 & $\num{267.3993} \frac{\diff{R}\diff{Y}}{\mean{Y}} + \num{13775.7831}m^2(1-m)^2(\diff{R})^2 + \num{7.1386}(1-m)\diff{Y}(\diff{R})^2 $ \\
& $ + \num{0.0135}$\\ \hline
4 & $\num{22.9437}\frac{\diff{R}\diff{Y}}{\mean{IP^{2+}}} + \num{13090.1088}m^2(1-m)^2(\diff{R})^2 + \num{5.9124}(1-m)\diff{Y}(\diff{R})^2 $ \\
& $ + \num{185.5363}\frac{(\diff{R})^2\diff{Y}}{\diff{M}} + \num{0.0040}$\\ \hline
5 & $\num{0.3198}m^2(1-m)^2\diff{Y} + \num{0.0001}\mean{IP^{3+}}\diff{IP3}\diff{Y} + \num{0.0335}\diff{R}\mean{IP^{3+}}\diff{Y} $\\[2pt]       
  & $\num{12977.1177}m^2(1-m)^2(\diff{R})^2 + \num{7.3732}(1-m)\diff{Y}(\diff{R})^2 - \num{0.0603}$\\  \hline
\end{tabular}

\vspace{10pt}

\begin{tabular}{|c|ccccc|} \hline
\diagbox{Error}{Desc. \#}&
\textrm{1}&
\textrm{2}&
\textrm{3}&
\textrm{4}&
\textrm{5}\\ \hline

MAE & 0.5581 & 0.1374 &  0.1249 &  0.1110 & 0.1075\\ \hline
MSE & 0.6803 & 0.0526 & 0.0403 & 0.0329 & 0.0260\\ \hline
ME & 4.0386 & 1.7908 & 1.5139 & 1.1327 & 1.2419\\ \hline
\end{tabular}
\label{S11}
\end{table}

In order to address the point (1)-(3) we apply the sparsification integrated with prior knowledge to a variety of scenarios: (i) use all modified descriptors illustrated in Sec.~\ref{sec:sparse-knowledge}, (ii) use all modified descriptors excluding the one based on ionic radii elementals $R^8$ and $R^9$, (iii) use all modified descriptors excluding the one based on the volume elemental $V$, and (iv) using only descriptors based on the volume $V$, mixing ratio $m$ and Young modulus Y elementals. In addition, we used three different set of data---monazite, xenotime, fused---for each of the four cases (i)--(iv). In total, we produced 12 separate sparsification scenarios, each specifying functions up to five leading terms and the relative errors.

\begin{table}[htbp]
\caption{Monazite: all elemental descriptors except Ionic radius coordination }
\begin{tabular}{|c|l|} \hline
\# & Functions\\ \hline
1 & $\num{0.9246}m(1-m)(\diff{V})^2 + \num{0.0173}$\\ \hline
2 & $\num{0.8044}m(1-m)(\diff{V})^2 + \num{0.5302}m^2(1-m)^2(\diff{V})^2 + \num{0.0220}$ \\ \hline
3 & $\num{0.8668}m(1-m)(\diff{V})^2 + \num{0.1626}m(1-m)^2(\diff{V})^2 - \num{0.00005}(\frac{\diff{V}}{m})^3 + \num{0.0108}$\\ \hline
4 & $\num{24.6538}\frac{\diff{V}}{\mean{IP^{3+}}\mean{Z_{eff}}} + \num{0.8378}m(1-m)(\diff{V})^2 + \num{0.1708}m(1-m)^2(\diff{V})^2 $ \\
& $- \num{0.00005} (\frac{\diff{V}}{m})^3 - \num{0.0337}$\\ \hline
5 & $\num{-0.0003}\frac{\diff{Y}}{m^3} + \num{40.2702}\frac{\diff{V}}{\mean{IP^{3+}}\mean{Z_{eff}}} + \num{0.8019}m(1-m)(\diff{V})^2 $ \\
& $+\num{0.2111}m(1-m)^2(\diff{V})^2 - \num{0.00003}(\frac{\diff{V}}{m})^3 - \num{0.0395}$ \\\hline
\end{tabular}

\vspace{10pt}

\begin{tabular}{|c|ccccc|} \hline
\diagbox{Error}{Desc. \#}& 
\textrm{1}&
\textrm{2}&
\textrm{3}&
\textrm{4}&
\textrm{5}\\ \hline
MAE & 0.0622 & 0.0575 & 0.0516 &  0.0500 & 0.0460\\ \hline
MSE & 0.0095 & 0.0082 & 0.0064 & 0.0058 & 0.0051\\ \hline
ME & 0.4936 & 0.6309 & 0.6195 & 0.5149 & 0.4814\\ \hline
\end{tabular}
\label{S2}
\end{table}

\begin{table}[htbp]
\caption{Monazite: all elemental descriptors except volume}
\begin{tabular}{|c|l|} \hline
\# & Functions\\ \hline
1 & $\num{19964.3970}m^2(1-m)^2(\diff{R})^2 + \num{0.0382}$\\  \hline
2 & $\num{12358.3498}m^2(1-m)^2(\diff{R})^2 + \num{113.7001}\frac{(\diff{R})^2\diff{Y}}{\diff{Z_{eff}}} - \num{0.0002}$ \\\hline
3 & $\num{11702.6552}m^2(1-m)^2(\diff{R})^2 + \num{116.2558}\frac{(\diff{R})^2\diff{Y}}{\diff{Z_{eff}}} + \num{6851.1011}\frac{(\diff{R})^2\diff{\chi}}{\diff{Y}} $\\
& $- \num{0.0278}$\\\hline
4 & $\num{0.1684}\frac{\mean{IP^{3+}}\mean{\chi}}{\mean{IP^{2+}}} + \num{11865.2535}m^2(1-m)^2(\diff{R})^2 + \num{113.8385}\frac{(\diff{R})^2\diff{Y}}{\diff{Z_{eff}}} $\\
& $+ \num{5585.8614}\frac{(\diff{R})^2\diff{\chi}}{\diff{Y}} - \num{0.3904}$\\\hline
5 & $\num{0.6122}m^2(1-m)^2\diff{Y} + \num{0.4085}\diff{IP^{3+}}\diff{\chi}\diff{IP^{2+}} + \num{34.8468}\frac{\diff{R}\diff{Y}}{\mean{IP^{2+}}} $\\[2pt]                   
& $ + \num{9966.2662}m^2(1-m)^2(\diff{R})^2 + \num{175670.7717}\frac{(\diff{R})^3}{\mean{IP^{2+}}\diff{Z_{eff}}} - \num{0.0642}$\\ \hline
\end{tabular}

\vspace{10pt}

\begin{tabular}{|c|ccccc|} \hline
\diagbox{Error}{Desc. \#}&
\textrm{1}&
\textrm{2}&
\textrm{3}&
\textrm{4}&
\textrm{5}\\ \hline

MAE & 0.2344 & 0.0858 & 0.0837 & 0.0807 & 0.0810\\ \hline
MSE & 0.1296 & 0.0176 & 0.0152 & 0.0145 & 0.0140\\ \hline
ME & 1.4019 & 0.7153 & 0.6658 & 0.6053 & 0.5780\\ \hline
\end{tabular}
\label{S3}
\end{table}

\begin{table}[htbp]
\caption{Xenotime: all elemental descriptors except Ionic radius coordination }
\begin{tabular}{|c|l|} \hline
\# & Functions\\ \hline

1 & $\num{1.2264}m(1-m)(\diff{V})^2 + \num{0.0630}$\\  \hline
2 & $\num{62658.0864}\frac{(\diff{V})^2}{(\mean{V})^3} + \num{2.5431}m^2(1-m)^2(\diff{V})^2 + \num{0.0094}$ \\ \hline
3 & $\num{53402.4559}\frac{(\diff{V})^2}{(\mean{V})^3}  + \num{2.5473}m^2(1-m)^2(\diff{V})^2 + \num{0.0003}(1-m)\mean{Y}(\diff{V})^2 $\\ 
& $+ \num{0.0123}$\\ \hline
4 & $\num{0.0432}(1-m)(\diff{V})^2 + \num{1524.3348}(\frac{\diff{V}}{\mean{V}})^2 - \num{1.3755}m(1-m)(\diff{V})^2 $\\ 
& $+ \num{5.7829}m^2(1-m)^2(\diff{V})^2 + \num{0.0047}$\\ \hline
5 & $\num{0.0430}(1-m)(\diff{V})^2 + \num{5499.6413}(\frac{\diff{V}}{\mean{V}})^2 - \num{200379.3047}\frac{(\diff{V})^2}{(\mean{V})^3} $\\[2pt]
  & $ - \num{3.5547}m(1-m)(\diff{V})^2 + \num{10.8335}m^2(1-m)^2(\diff{V})^2 + \num{0.0114}$\\ \hline
\end{tabular}

\vspace{10pt}

\begin{tabular}{|c|ccccc|} \hline
\diagbox{Error}{Desc. \#}&
\textrm{1}&
\textrm{2}&
\textrm{3}&
\textrm{4}&
\textrm{5}\\ \hline

MAE & 0.1699 & 0.0817 &  0.0492 &  0.0396 & 0.0348\\ \hline
MSE & 0.0688 & 0.0243 & 0.0058 & 0.0033 & 0.0024\\ \hline
ME & 1.1805 & 0.9838 & 0.4178 & 0.3495 & 0.3412\\ \hline
\end{tabular}
\label{S6}
\end{table}

\begin{table}[htbp]
\caption{Xenotime: all elemental descriptors except volume}
\begin{tabular}{|c|l|} \hline
\# & Functions\\ \hline

1 & $\num{9.8282}\frac{\diff{Y}\diff{Z_{eff}}}{\mean{Y}} - \num{0.0317}$\\  \hline
2 & $\num{18347.9196}m^2(1-m)^2(\diff{R})^2 + \num{3136.8386}\frac{(\diff{R})^2\mean{IP^{3+}}}{\mean{Y}} - \num{0.0109}$ \\ \hline
3 & $\num{18059.4867}m^2(1-m)^2(\diff{R})^2 + \num{2625.5857}\frac{(\diff{R})^2\mean{IP^{3+}}}{\mean{Y}}  $\\ 
& $ + \num{93534.5878}(1-m)\frac{(\diff{R})^3}{\mean{IP^{2+}}} + \num{0.0577}$\\ \hline
4 & $\num{0.3421}\frac{\diff{Z_{eff}}\diff{Y}}{\mean{IP^{2+}}} + \num{17191.3100}m^2(1-m)^2(\diff{R})^2 + \num{12680655.2904}\frac{(\diff{R})^3}{\mean{Y}\diff{M}} $\\
& $ + \num{89.9428}(1-m)(\diff{R})^3\mean{IP^{3+}} + \num{0.0066}$\\ \hline
5 & $\num{0.3441}\frac{\diff{Z_{eff}}\diff{Y}}{\mean{IP^{2+}}} + \num{6e-05}\mean{IP^{3+}}\diff{IP^{3+}}\diff{Y} + \num{17407.1}m^2(1-m)^2(\diff{R})^2 $\\[2pt]                     & $+ \num{10464580.7}\frac{(\diff{R})^3}{\mean{Y}\diff{M}} + \num{91.8}(1-m)(\diff{R})^3\mean{IP^{3+}} + \num{0.0054}$ \\ \hline
\end{tabular}

\vspace{10pt}

\begin{tabular}{|c|ccccc|} \hline
\diagbox{Error}{Desc. \#}&
\textrm{1}&
\textrm{2}&
\textrm{3}&
\textrm{4}&
\textrm{5}\\ \hline

MAE & 0.3575 & 0.1547 & 0.1321 & 0.0871 & 0.0813\\ \hline
MSE & 0.3691 & 0.0603 &  0.0335 & 0.0162 & 0.0139\\ \hline
ME & 3.0150 & 1.5249 & 0.7334 & 0.6216 & 0.6191\\ \hline
\end{tabular}
\label{S7}
\end{table}

The (1) statement is inferred directly from the direct inspection of the tables for cases (ii) and (iii) applied to the fused data set (Tables \ref{S10} and \ref{S11}). From these two tables, one observes that the functions of one up to five leading terms are exactly the same except that wherever in one table the ionic radius $R$ appears in the other the volume $V$ appears in exactly the same terms. Moreover, all of the coefficients of the leading terms are exactly the same across the two tables as well as the all the errors. This obviously points out to the fact that only either ionic radius or volume should be included in the list of elemental properties. In order to understand which among these two elemental properties should be eliminated, we look next to the cases (ii) and (iii) applied separately to monazite and xenotime data sets (Tables \ref{S2}, \ref{S3}, \ref{S6}, and \ref{S7}). Comparing the errors, it is immediately obvious that descriptors without the volume $V$ elemental returns always larger error than descriptors without the ionic radius $R$ elemental for both monazite and xenotime data sets. This is a strong indication that $V$ should be preferred over $R$ as the elemental property of choice. This concludes statement (1) and excludes from further analysis cases (i) and (iii).    

\begin{figure}[ht!]
    \centering
    \begin{subfigure}{0.49\textwidth}
        \centering
        \includegraphics[width=\textwidth]{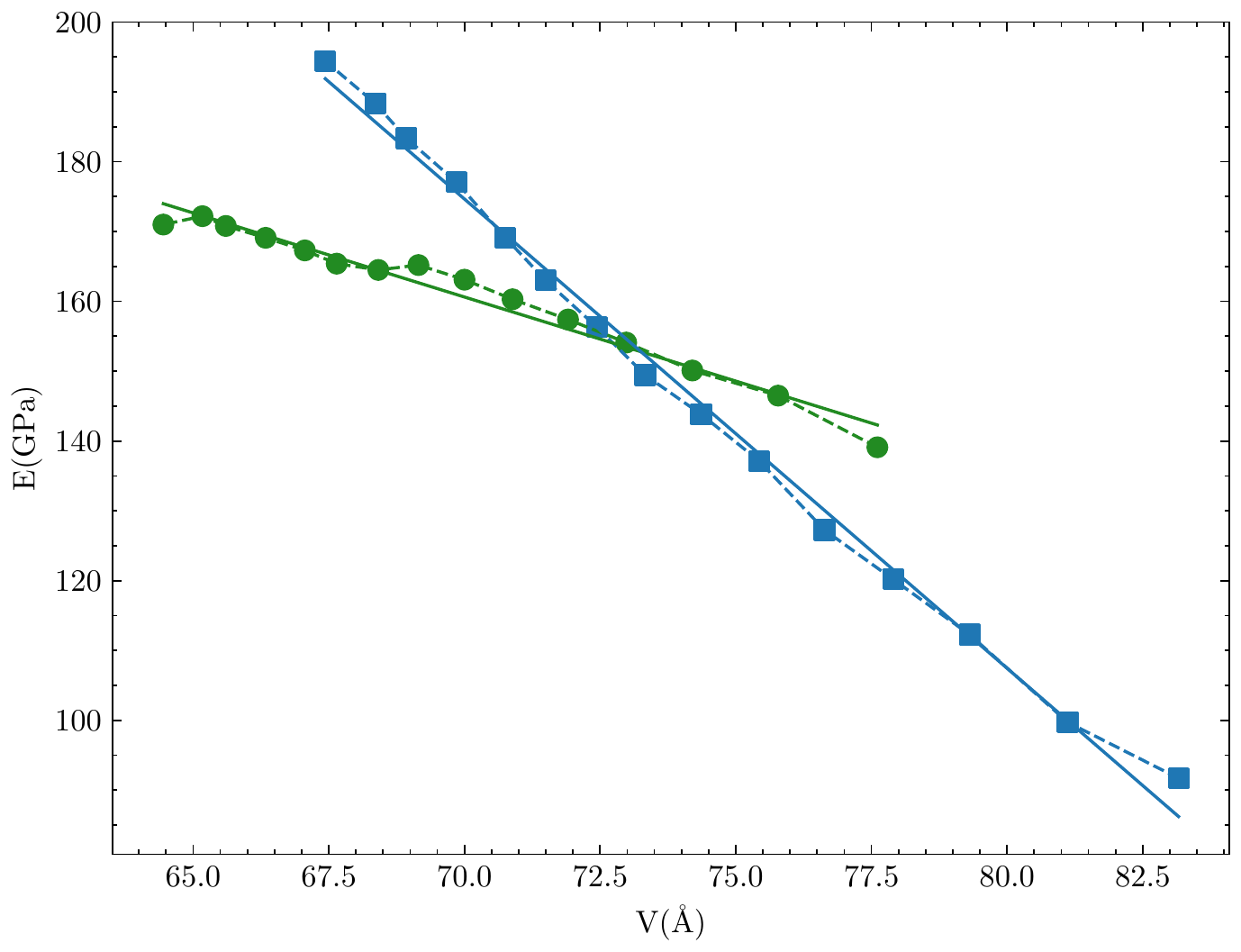} 
        \caption{The Young's modulus as a function of volume for monazite (green circles) and xenotime (blue squares) phases. The lines represent the linear regression fits. 
        \label{EV}}
    \end{subfigure}
    \hfill
    \begin{subfigure}{0.49\textwidth}
        \centering
        \includegraphics[width=\textwidth]{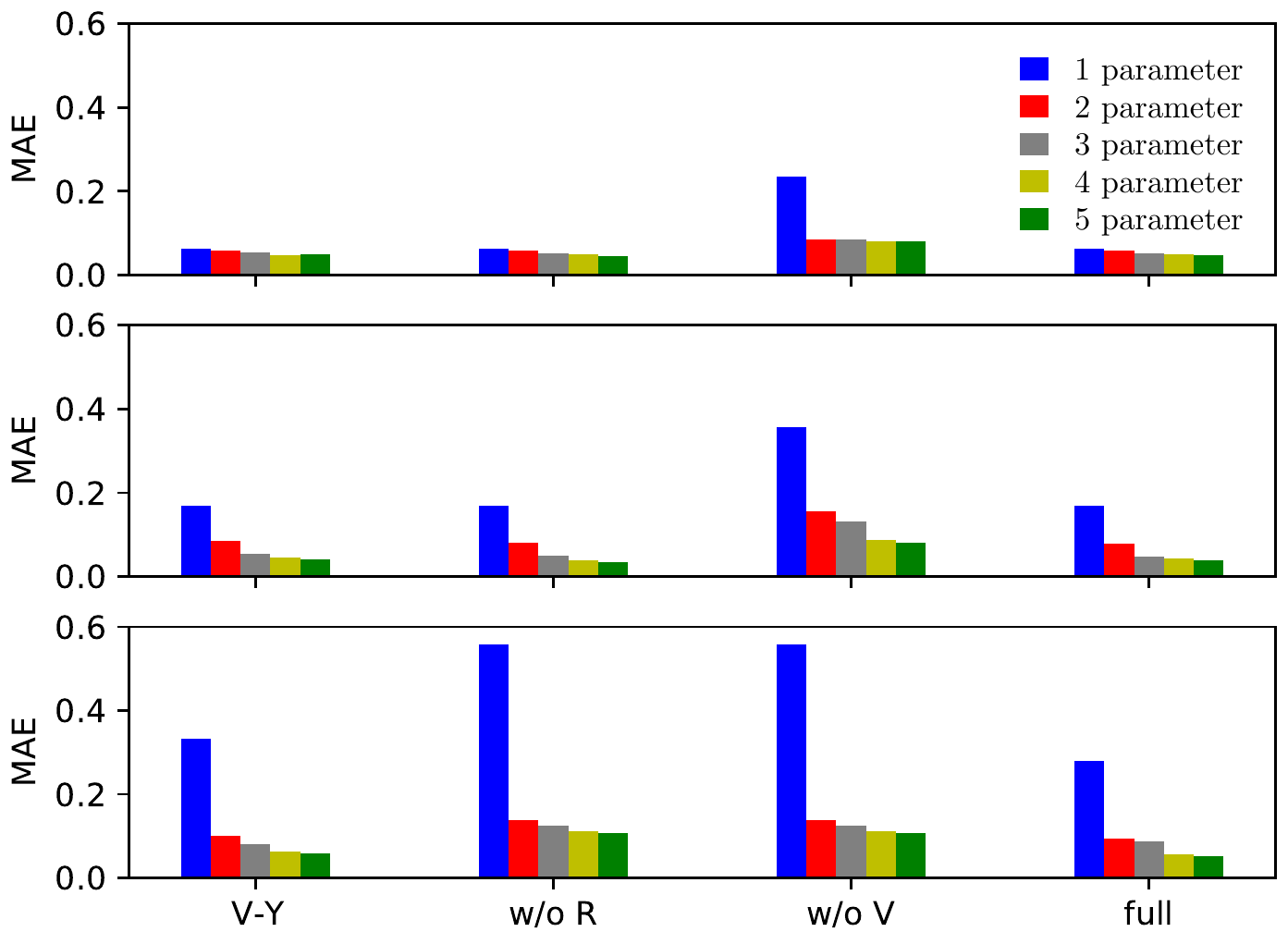}
        \caption{The mean absolute error in kJ/mol of the LASSO +
          $\ell_0$ sparsification for the monazite, xenotime and fused
          cases obtained with the set of 1 to 5 leading terms. \label{fmae}}
    \end{subfigure}
    \caption{Young modulus vs Volume and MAE for the LASSO + $\ell_0$ sparsification.}  
\end{figure}

Next, we look at cases (iii) and (iv) applied to all three sets of data (Tables \ref{S1}, \ref{S2}, \ref{S5}, \ref{S6}, \ref{S9}, and \ref{S10}). The very first observation is that no matter what data set or case one looks at, all the one-term function are of the form $m(1-m)(\Delta V)^2$. Unequivocally this is the first leading term describing the excess of enthalpy $H^E$. We also notice that sub-leading terms  for monazite and xenotime data sets are dominated by the volume elemental but this dominance manifests itself differently for each distinct data set and with errors that vary from case to case. In the case of xenotime, the prediction of $H^E$ with only one term is not sufficient and additional terms are needed to get satisfactory accuracy. While the leading term is also $m(1-m)\Delta V^2$, the prefactor coefficient is substantially larger that the one obtained for monazite (1.23 vs. 0.94). Although this difference could be possibly explained with on average larger Young's moduli of xenotime phases than monazite (see Eq. \ref{eq:E1}), this demonstrates that additional terms are essential to obtain an equation that consistently described both structures. When we look at the solutions for xenotime alone, different terms appear, usually containing some power of the difference and average of volume and/or Young's modulus. Only simultaneous fit to monazite and xenotime data sets could shed a light on the composition of the second leading term.

\begin{table}[htbp]
\caption{Monazite: only Volume and Young modulus}
\begin{tabular}{|c|l|}
\hline
\# & Functions\\ \hline
1 & $\num{0.9247}m(1-m)(\diff{V})^2 + \num{0.0173}$\\ \hline
2 & $\num{0.3573}m(1-m)\diff{V} + \num{0.8674}m(1-m)(\diff{V})^2 - \num{0.066}$\\ \hline
3 & $\num{0.00004}\frac{(\diff{Y})^2}{m^3} + \num{0.8681}m(1-m)(\diff{V})^2 +\num{ 0.1527}m(1-m)^2(\diff{V})^2 +\num{0.0207}$\\ \hline
4 & $\num{0.1028}m(1-m)\diff{Y} - \num{0.00004}\frac{(\diff{Y})^2}{m^3} +\num{0.8243}m(1-m)(\diff{V})^2 $ \\
& $+ \num{ 0.1574}m(1-m)^2(\diff{V})^2 -\num{0.0299}$\\ \hline
5 & $\num{0.1204}m(1-m)\diff{Y} - \num{0.00007}\frac{(\diff{Y})^2}{m^3} + \num{0.9109}m(1-m)(\diff{V})^2$\\[2pt]
  & $+\num{0.2370}m(1-m)^2(\diff{V})^2 -\num{0.5384}m^2(1-m)^2(\diff{V})^2 - \num{0.0413}$ \\\hline
\end{tabular}

\vspace{10pt}

\begin{tabular}{|c|ccccc|}
\hline
\diagbox{Error}{Desc. \#}&
\textrm{1}&
\textrm{2}&
\textrm{3}&
\textrm{4}&
\textrm{5}\\
\hline
MAE & 0.0622 & 0.0585 & 0.0546 &  0.0482 & 0.0495\\
MSE & 0.0095 & 0.0082 & 0.0068 & 0.0057 & 0.0053\\
ME & 0.4936 & 0.4239 & 0.6279 & 0.5357 & 0.4579\\ \hline
\end{tabular}
\label{S1}
\end{table}

\begin{table}[htbp]
\caption{Xenotime: only Volume and Young modulus}
\begin{tabular}{|c|l|} \hline
\# & Functions\\ \hline

1 & $\num{1.2264}m(1-m)(\diff{V})^2 + \num{0.0630}$\\  \hline
2 & $\num{0.0010}\mean{Y}(\diff{V})^2 + \num{2.5799}m^2(1-m)^2(\diff{V})^2 + \num{0.0296}$ \\ \hline
3 & $\num{0.0008}\mean{Y}(\diff{V})^2 + \num{2.5799}m^2(1-m)^2(\diff{V})^2 + \num{0.0003}(1-m)\mean{Y}(\diff{V})^2 $\\
& $ + \num{0.0296}$\\ \hline
4 & $ \num{-0.000008}(\mean{Y})^2(\diff{V})^2 + \num{2.5772}m^2(1-m)^2(\diff{V})^2 + \num{0.0003}(1-m)\mean{Y}(\diff{V})^2 $\\
& $ + \num{11.8681}\frac{\mean{Y}(\diff{V})^2}{(\mean{V})^2}+ \num{0.0031}$\\ \hline
5 & $\num{-0.00002}(\mean{Y})^2(\diff{V})^2 - \num{0.5726}m(1-m)(\diff{V})^2+ \num{3.9222}m^2(1-m)^2(\diff{V})^2 $\\[2pt] 
  & $ + \num{0.0003}(1-m)\mean{Y}(\diff{V})^2 + \num{23.1551}\frac{\mean{Y}(\diff{V})^2}{(\mean{V})^2} - \num{0.0154}$\\ \hline
\end{tabular}

\vspace{10pt}

\begin{tabular}{|c|ccccc|} \hline
\diagbox{Error}{Desc. \#}& 
\textrm{1}&
\textrm{2}&
\textrm{3}&
\textrm{4}&
\textrm{5}\\ \hline
MAE & 0.1699 & 0.0850 & 0.0547 &  0.0449 & 0.0408\\ \hline
MSE & 0.0688 & 0.0252 & 0.0067 & 0.0046 & 0.0037\\ \hline
ME & 1.1805 & 1.0853 & 0.5429 & 0.3859 & 0.3759\\ \hline
\end{tabular}
\label{S5}
\end{table}


Looking at the fused data, it becomes clear that not only the volume but also the Young modulus $Y$ elemental plays a central role in sub-leading terms. In addition, comparing Tables \ref{S9} and \ref{S10}, it is immediately clear that using descriptors of just $V$ and $Y$ returns consistently smaller errors for all sub-leading terms. Incidentally, 
Young's moduli and volumes/ionic radii are the elemental properties used in models of \cite{M07} and \cite{KL16}. 
From this analysis we infer that we should focus on the leading and sub-leading terms provided by table \ref{S9}, where all elementals apart from $V$ and $Y$ are considered. 
The leading second term contains product of $\Delta V$ and $\Delta Y$. With combination of $m(1-m)\Delta V^2$ and $\Delta Y \Delta V / \Bar{V}^2$ one can satisfactorily express $H^E$ for both sets of data. 
This indicates that in addition to difference in volume, the difference in Young's modulus plays a significant role in determining the excess enthalpy of mixing. To better understand the role of this second term, we plot in Fig. \ref{EV} the variation of Young's modulus as a function of $\Delta V$ for the two phases. For the considered range of elastic moduli and volumes, there is a clear linear relationship between the two. However, the linear relationship is much steeper in case of xenotime than monazite, with the slope about three times higher in the former case. This makes the $\Delta Y \Delta V / \Bar{V}^2$ term much larger (by a factor $\sim 3$) in case of xenotime, putting it on equal footing with the leading term. This explains why missing this term, there is a factor of $\sim 2$ difference between the {\it ab initio} data and the discussed, prior theoretical models (see Fig.~\ref{mon-xen}).

In general, adding more terms improves the fit only marginally (see
Fig. \ref{fmae}). On the other hand, one can observe from table
\ref{S9} that for any choice of number of terms, the factor $\Delta Y
\Delta V / \Bar{V}^2$ is always present, confirming its importance in
contributing to the expression of $H^E$. This concludes the statements
(2) and (3).

\begin{table}[htbp]
\caption{Monazite+Xenotime: only Volume and Young modulus}
\begin{tabular}{|c|l|}
\hline
\# & Functions\\ \hline
1 & $\num{1.1453}m(1-m)(\diff{V})^2 - \num{0.0517}$\\ \hline 
2 & $\num{2.9453}m^2(1-m)^2(\diff{V})^2 + \num{108.1079}\frac{\diff{V}\diff{Y}}{(\mean{V})^2}+ \num{0.0429}$ \\ \hline
3 & $\num{0.0066}m(1-m)^2(\diff{Y})^2 + \num{0.7963}m(1-m)(\diff{V})^2 + \num{54.3783}\frac{\diff{Y}\diff{V}}{(\mean{V})^2} $\\
& $ + \num{0.0064}$\\ \hline
4 & $\num{0.0047}m(1-m)^2(\diff{Y})^2 + \num{0.4672}m(1-m)(\diff{V})^2 + \num{1.2743}m^2(1-m)^2(\diff{V})^2 $\\
& $ + \num{70.8755}\frac{\diff{V}\diff{Y}}{(\mean{V})^2} + \num{0.0195}$\\ \hline
5 & $\num{0.5117}m(1-m)\diff{V} + \num{0.0001}(1-m)\diff{V}(\diff{Y})^2 + \num{0.2862}m(1-m)(\diff{V})^2 $\\[2pt] 
  & $ + \num{1.639}m^2(1-m)^2(\diff{V})^2 + \num{78.1966}\frac{\diff{V}\diff{Y}}{(\mean{V})^2} - \num{0.0764}$  \\\hline
\end{tabular}

\vspace{10pt}

\begin{tabular}{|c|ccccc|} \hline
\diagbox{Error}{Desc. \#}& 
\textrm{1}&
\textrm{2}&
\textrm{3}&
\textrm{4}&
\textrm{5}\\ \hline

MAE & 0.3316 & 0.0997 & 0.0801 &  0.0630 & 0.0594\\ \hline
MSE & 0.2804 & 0.0262 & 0.0173 & 0.0091 & 0.0083\\ \hline
ME & 2.8122 & 1.3163 & 0.7561 & 0.6165 & 0.5345\\ \hline
\end{tabular}
\label{S9}
\end{table}



\section{Summary and conclusions}
\label{sec:conclusion}

In this work we report on a 3-step approach that combines distinct methods from classical Machine Learning to reach a {\it scientifically consistent explainable} methodology. First, we use a KRR approach on the generated data and evaluate which kernel returns the least amount of error. Then, using the results of KRR, we reverse engineer the model and proceed to sparsify the vector of coefficients using a so-called LASSO$+\ell_0$ procedure. Finally we integrate domain-specific knowledge to force an a-posteriori scientific consistency of the reverse model.

In order to demonstrated the feasibility and potential of this methodology, we have applied it to the computation of the excess of enthalpy of formation $H^E$ of solid solutions of lanthanide Orthophosphates. This particular class of materials is present in nature in two distinct crystal structure---monazite and xenotime---for which no single formula is capable of describing accurately $H^E$. Applying our machine learning based 3-step method to a set of in silico data, we were able to retrieve sub-leading corrections to known expressions for $H^E$, which represent an important step in resolving a conflicting description of the excess of enthalpy for both type of structures. We expect that the importance of accounting for the gradient of elastic moduli when estimating the excess enthalpy of mixing will trigger follow-up theoretical studies aiming at providing physical interpretation of the origin of this phenomenon.
The successful application of our procedure shows the potential of its application to other areas of Quantum Chemistry and Materials Science where explainability of machine learning models is an essential feature.


\section*{Acknowledgments}
The authors gratefully acknowledge the computing time
granted by the JARA Vergabegremium and provided on
the JARA Partition of the supercomputers JURECA
at Forschungszentrum Jülich and CLAIX at RWTH Aachen
University.

\bibliographystyle{siamplain}
\bibliography{literature/collection,literature/m}

\end{document}